\documentclass[a4paper,fleqn]{cas-dc}

\usepackage[numbers]{natbib}

\usepackage{amsmath}
\usepackage{amssymb}
\usepackage{amsfonts}
\usepackage{bm}
\usepackage{times,float}
\usepackage{graphicx}
\usepackage[dvipsnames,svgnames]{xcolor}
\usepackage{hyperref}
\hypersetup{colorlinks=true, linkcolor=NavyBlue, citecolor=PineGreen,urlcolor=cyan}
\usepackage{physics}
\graphicspath{{FIGURES/}}
\usepackage{subcaption}
\usepackage{cuted}

\newcommand{\bmp}[1]{\bm{#1}^{\prime}}
\newcommand{\partd}[3][1]{%
  \ifnum#1=1
    \frac{\partial #2}{\partial #3}%
  \else
    \frac{\partial^{#1}#2}{\partial #3^{#1}}%
  \fi
}
\newcommand{\parenth}[1]{\left( #1 \right) }
\newcommand{\parenthc}[1]{\left[ #1 \right] }

\def\tsc#1{\csdef{#1}{\textsc{\lowercase{#1}}\xspace}}
\tsc{WGM}
\tsc{QE}

\begin{document}
\let\WriteBookmarks\relax
\def\floatpagepagefraction{1}
\def\textpagefraction{.001}

\shorttitle{}    

\shortauthors{}  

\title [mode = title]{Charge and energy transport in graphene with smooth finite-range disorder}  




\author{Juan A. Ca\~nas}[orcid = 0009-0007-5126-7281]

\cormark[1]


\ead{juan.canas@correo.nucleares.unam.mx}


\credit{Conceptualization, Formal analysis, Methodology, Writing – original draft}

\affiliation{organization={Instituto de Ciencias Nucleares, Universidad Nacional Aut\'{o}noma de M\'{e}xico},
%          citysep={}, % Uncomment if no comma needed between city and postcode
            postcode={04510}, 
            state={Ciudad de M\'{e}xico},
            country={M\'exico}}

\author{Daniel A. Bonilla}[orcid = 0000-0002-9719-5371]

% Footnote of the second author

% Email id of the second author
\ead{daniel.bonillam@correo.nucleares.unam.mx}


\credit{Methodology, Visualization, Writing – review \& editing}


\author{J. C. Pérez-Pedraza}[orcid = 0000-0002-7673-4778]

% Footnote of the third author

% Email id of the second author
\ead{julio.perez@correo.nucleares.unam.mx}


\credit{Conceptualization, Writing – review \& editing}

\author{A. Mart\'{i}n-Ruiz}[orcid = 0000-0001-5308-5448]

% Footnote of the fourth author

% Email id of the second author
\ead{alberto.martin@nucleares.unam.mx}


\credit{Supervision, Writing – review \& editing}

\cortext[1]{Corresponding author}



\begin{abstract}
We investigate charge and energy transport in monolayer graphene with smooth finite-range disorder, modeled by soft impurity potentials. Using a continuum Dirac model, we go beyond the Born approximation by computing the exact scattering matrix for individual impurities. This captures the full nonperturbative physics of smooth disorder. From the exact scattering data, we evaluate transport coefficients by solving the Boltzmann equation with energy-resolved phase shifts. We analyze electrical and electronic thermal conductivities versus carrier density and temperature, including deviations from the Wiedemann-Franz law. Our results reveal that finite-range disorder nontrivially modifies charge and heat currents, especially at low energies where perturbative methods fail. These findings provide a more accurate transport characterization for disordered Dirac materials and clarify how smooth disorder governs energy flow in graphene.
\end{abstract}


%\item 

%\nocite{*}

% Keywords

\maketitle

% Main text

\section{Introduction}

The isolation of monolayer graphene inaugurated a new platform for studying massless Dirac fermions and quantum transport in two dimensions, an achievement recognized with the 2010 Nobel Prize in Physics \cite{NovoselovScience2004,NovoselovNature2005}. Beyond its fundamental interest, graphene's exceptional carrier mobility and record-high lattice thermal conductivity make it a prime testbed for charge, heat, and coupled thermoelectric transport \cite{CastroNetoRMP2009,DasSarmaRMP2011,BalandinNatMat2011}. Thermopower and Nernst measurements have mapped energy-dependent scattering and electron-hole asymmetries across wide density and temperature windows, highlighting both the opportunities for thermoelectric control and the sensitivity of transport to microscopic scattering \cite{ZuevPRL2009,WeiPRL2009}. More recently, graphene-based systems have exhibited an exceptionally rich physical phenomenology, including electrically tunable band gaps and correlated electronic states in bilayer graphene \cite{McCannPRL2006}, as well as flat-band physics, unconventional superconductivity, and strongly correlated phases in magic-angle twisted bilayer graphene \cite{CaoNature2018SC,CaoNature2018Correlated}. Alongside these developments, a number of transport studies in bilayer graphene \cite{OptMat2016,EPJB2015} and other graphene based materials \cite{IJMPB2015, OQE2024,ApplPhysA2024a,ApplPhysA2025}  have demonstrated that different impurity types and disorder profiles can strongly influence electronic and optical responses. Although bilayer and twisted systems host distinct band structures and symmetry properties compared to monolayer graphene, these works collectively underscore the central role of disorder and scattering mechanisms in shaping charge and energy transport across graphene-based and Dirac materials.

In realistic devices, the dominant disorder component is often characterized by smooth, finite-range potentials, arising from remote charged impurities, interface roughness, or gentle strain, thereby suppressing intervalley processes while reshaping low-energy scattering and current flow \cite{AdamPNAS2007,HwangPRL2007,MartinNatPhys2008,ChenNatPhys2008}. These features place a premium on energy-resolved descriptions of how specific disorder profiles govern both charge and heat transport in the low-energy regime. In this context, smooth finite-range scatterers may be viewed as an effective description of local smooth disorder, such as long-wavelength surface roughness or substrate-induced potential fluctuations, provided the perturbation varies on length scales larger than the lattice spacing and intervalley scattering remains suppressed. Isolated defects in graphene are frequently modeled as short-range perturbations, approximating them as point-like scatterers \cite{HwangPRL2007,AdamPE2008,PhysRevB.91.115410}. However, this model requires careful physical interpretation as it implicitly introduces an effective range, which emulates the role of a scattering cross-section \cite{Ferry2013}. Furthermore, the physical structure of defects can be spatially extended, leading to potential barrier-like effects \cite{Koepke2013}. A natural generalization, therefore, is to replace the short-range potential model with circular, finite-radius soft scatterers, defined by a radius $R$ and a constant strength $V_0$, which we will refer to as \textit{soft-spheres}. This formulation self-consistently accounts for both the effective scattering range and the physical extension of the impurity, and aligns with established graphene electron-optics/Mie-like scattering setups \cite{WuPRB2014,CaridadNatComm2016}. Within this phenomenological framework, soft-sphere potentials can also capture the effect of screened charged impurities in the substrate, with the radius $R$ encoding an effective screening length and spatial extension of the perturbation, without requiring a detailed microscopic description of the charge distribution.

A large body of theory models disorder-limited transport in graphene within the Born or self-consistent Born approximation (SCBA), which is convenient but becomes unreliable near charge neutrality and for strong or spatially extended scatterers typical of smooth finite-range disorder \cite{PeresRMP2010,ShonAndoJPSJ1998,KlosPRB2010}. Renormalization-group and nonperturbative analyses show that such scatterers can drive logarithmic flows and resonant/unitary-limit behavior that lie outside perturbation theory, qualitatively altering low-energy cross sections and conductivity scaling \cite{AleinerEfetovPRL2006,OstrovskyPRB2006,RobinsonPRL2008,WehlingPRL2010}. While exact scattering and phase-shift frameworks have clarified aspects of charge transport for particular impurity classes \cite{PhysRevB.91.115410,NovikovPRB2007,FerreiraPRB2011}, a unified, energy-resolved nonperturbative treatment that consistently links charge and heat currents for smooth finite-range disorder remains scarce. 

In this work, we consider a dilute ensemble of soft-spheres with radii much larger than the lattice spacing. Electron-electron and electron-phonon interactions are neglected within the experimentally relevant window in which elastic impurity scattering dominates, as is typical for moderate temperatures and high-quality encapsulated devices. Methodologically, we compute the exact single-impurity $S$-matrix for a circular soft potential, extract the partial-wave phase shifts, and use them to obtain the energy-resolved transport scattering time. These quantities feed into a linearized Boltzmann analysis of charge and heat transport. In the DC and dilute-impurity limit considered here, the Boltzmann formulation based on exact phase shifts is fully equivalent to the Kubo approach, while offering a more transparent connection between finite-range scattering and thermoelectric transport. \cite{NovikovPRB2007,FerreiraPRB2011,WuPRB2014}.

Our results demonstrate that these soft scatterers impurities act as non-resonant scattering centers, fundamentally shaping the transport behavior. Our central finding is that the spatial extent (radius) of the defects, rather than their potential strength, emerges as the dominant parameter controlling key thermoelectric quantities, including the violation of the Wiedemann-Franz law, the Seebeck coefficient, and the figure of merit, $ZT$. While $ZT$ is fundamentally defined by the total thermal conductivity, our analysis focuses on the electronic component to isolate the impact of smooth disorder on carrier transport. These results identify defect size as a crucial tuning parameter and suggest that combining nanoscale impurity engineering with strategies to reduce phononic thermal conductivity provides a synergistic route toward designing high-efficiency thermoelectric graphene.

The remainder of this paper is organized as follows. Section \ref{model} introduces pristine monolayer graphene and reviews the key ingredients of the low-energy model and notation used throughout. Section \ref{phase_shifts} solves the Dirac-like scattering problem for circular soft scatterers and derives exact partial-wave phase shifts analytically. Section \ref{transport_section} develops the semiclassical transport framework based on energy-resolved phase shifts, leading to expressions for the electrical conductivity, the electronic thermal conductivity, the Lorenz coefficient, the Seebeck coefficient, and the thermoelectric figure of merit. Section \ref{results_discussion_section} reports and analyzes the numerical results across carrier density and temperature ranges, emphasizing regimes where smooth finite-strength disorder produces departures from standard expectations. Section \ref{conclusion_section} concludes with a summary and an outlook on extensions and experimental implications.

\section{Basics of the model} \label{model}
Graphene is composed of carbon atoms forming a two-dimensional honeycomb structure that can be described as a hexagonal arrangement of two inequivalent triangular sub-lattices, commonly labeled $A$ and $B$. The nearest-neighbor distance between carbon atoms is $a \approx 0.142$ nm, which establishes the characteristic microscopic length scale of the crystal \cite{CastroNetoRMP2009}. The underlying Bravais lattice is spanned by the primitive vectors
\begin{align}
    \bm{a}_{1} &= \left( \frac{\sqrt{3}a}{2}, \, \frac{3a}{2} \right),     &    \bm{a}_{2} &= \left( \frac{\sqrt{3}a}{2}, \, - \frac{3a}{2} \right),
\end{align}
so that the positions of $A$-sublattice sites are $\bm{R}_n = n_1 \bm{a}_1 + n_2 \bm{a}_2$, with integers $n_1,n_2$. Each atom in sub-lattice $A$ has three nearest neighbors in sub-lattice $B$, connected by
\begin{equation}
    \bm{\rho}_{1} = (0,a), 
    \bm{\rho}_{2} = \left( \frac{\sqrt{3}a}{2}, -\frac{a}{2} \right), 
    \bm{\rho}_{3} = \left( -\frac{\sqrt{3}a}{2}, -\frac{a}{2} \right).
\end{equation}

The tight-binding Hamiltonian with only nearest-neighbor hopping is
\begin{align}
    \hat{\mathcal{H}} = -t \sum_{\bm{R}_n}\sum_{j=1}^3 
    \hat{a}^{\dagger}_{\bm{R}_n}\hat{b}_{\bm{R}_n+\bm{\rho}_j} + \text{H.c.},
\end{align}
where $t$ is the nearest-neighbor hopping amplitude, experimentally and theoretically estimated as $t \approx 2.7$ eV~\cite{T_Reich_2002}. The operators $\hat{a}^{\dagger}_{\bm{R}_n}$ and $\hat{b}_{\bm{R}_n+\bm{\rho}_j}$ create and annihilate electrons on the two sub-lattices. After Fourier transforming, the Hamiltonian can be written in momentum space as
\begin{align}
    \hat{\mathcal{H}} = \sum_{\bm{k}}
    \begin{pmatrix}
    \hat{a}^{\dagger}_{\bm{k}} & \hat{b}^{\dagger}_{\bm{k}}
    \end{pmatrix}
    \begin{pmatrix}
    0 & f_{\bm{k}} \\
    f_{\bm{k}}^{*} & 0
    \end{pmatrix}
    \begin{pmatrix}
    \hat{a}_{\bm{k}} \\
    \hat{b}_{\bm{k}}
    \end{pmatrix},
\end{align}
with the structure factor
\begin{align}
    f_{\bm{k}} = -t\sum_{j=1}^3 e^{i\bm{k}\cdot \bm{\rho}_j}
\end{align}

The band structure is given by $E_{\pm}(\bm{k}) = \pm |f_{\bm{k}}|$, which reveals conical band crossings at six points in the Brillouin zone. Only two of these are inequivalent, the so-called Dirac points $\bm{K}_D^{\pm} = (\pm 4\pi/3\sqrt{3}a, 0)$, located at the zone corners~\cite{Wallace1947}. These valleys are related by time-reversal symmetry and host low-energy quasiparticles that behave as relativistic fermions.

Expanding around $\bm{K}_D^{\chi}$ with valley index $\chi=\pm 1$, writing $\bm{k} = \bm{K}_D^{\chi} + \bm{q}$ and assuming $|\bm{q}|\ll |\bm{K}_D^{\chi}|$, one finds
\begin{align}
    f_{\bm{K}_D^{\chi}+\bm{q}} \simeq \chi \hbar v_F (q_x - i \chi q_y),
\end{align}
with Fermi velocity $v_F = \frac{3at}{2\hbar} \approx 9.8 \times 10^{5} \, \text{m/s}$. This approximation turns the tight-binding Hamiltonian into a continuum description of massless Dirac fermions,
\begin{align}
    \hat{\mathcal{H}}_{0\chi}(\bm{p}) = \chi v_F \, \hat{\bm{\sigma}}\cdot \bm{p},
    \label{eq:PGraph_Ham}
\end{align}
where $\bm{p} = \hbar \bm{q}$ is the momentum measured relative to the Dirac point and $\hat{\bm{\sigma}} = (\sigma_x, \sigma_y)$ acts in the sub-lattice (pseudospin) space. The eigenvalues are
\begin{align}
    \mathscr{E}_{s\chi}(\bm{q}) = s \chi \hbar v_F q,
    \label{eq:Disp_Rel}
\end{align}
with $s=\pm 1$ labeling conduction and valence bands. This compact Dirac form emphasizes that low-energy carriers in graphene behave as chiral, massless fermions, a fact that underpins its unusual transport phenomena \cite{CastroNetoRMP2009}.

In realistic graphene samples, typical sources of disorder, such as charged impurities in the substrate, corrugations, and adsorbates, generate potentials that vary smoothly on the scale of the lattice constant. For this reason, modeling disorder with finite-range potentials is not merely a simplification but an accurate physical description \cite{CastroNetoRMP2009,DasSarmaRMP2011,HwangPRL2007,PeresRMP2010}. Such potentials naturally suppress intervalley scattering, favor small-angle processes, and can host nonperturbative resonances that dominate low-energy transport \cite{CastroNetoRMP2009,HwangPRL2007,PeresRMP2010}.

A common minimal model treats isolated defects as zero-range (delta-function) scatterers \cite{HwangPRL2007,AdamPE2008}. However, to properly account for the finite spatial extent and smooth profile of realistic defects, this approach must be generalized. In this work, we adopt a paradigmatic finite-range model: scalar impurities are represented as circular potentials of constant strength $V_0$ and radius $R$, refereed to as soft spheres. This model interpolates between the short-range limit $(R\rightarrow 0)$ and the long-range, barrier-like regime. Moreover, this model can be physically motivated by considering charged impurities in the substrate \cite{HwangPRL2007}, which, due to the screening, produces a short-range potential that, when projected onto the graphene plane and truncated at a distance controlled by the graphene-substrate separation and screening length, resembles a smooth, finite-range bump. This results in a smooth, finite-range perturbation that, to a good approximation, can be described with the soft sphere model, with $V_0$ and $R$ effectively encoding the charge, screening length and graphene-substrate distance information. Independently of the microscopic origin of the screened Coulomb-type interaction between the substrate-graphene layers\cite{AdamPNAS2007,HwangPRL2007}, the soft-sphere model we use in this work captures this essential phenomenology, while enabling an exact nonperturbative solution and capturing essential scattering physics which modulates the electronic transport \cite{NovikovPRB2007}.

\begin{figure}
    \centering
    \includegraphics[width=0.7\linewidth]{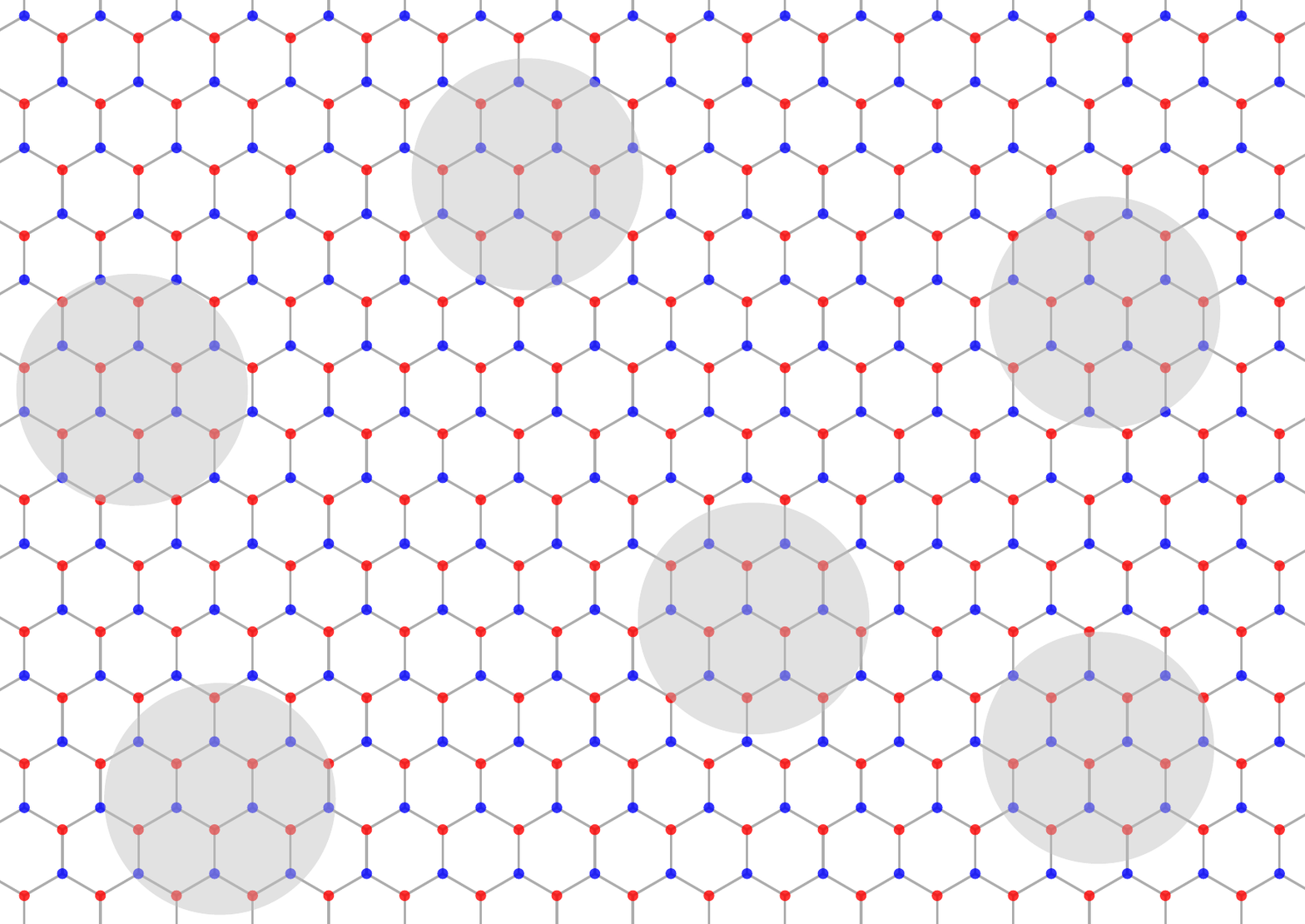}
    \caption{A schematic of the proposed model. Soft-sphere scatterer potentials are randomly distributed across a graphene sheet. The dilute impurity limit ensures statistically independent scattering centers with negligible overlap.}
    \label{fig:Model}
\end{figure}

We therefore consider a dilute distribution of soft spheres. The condition $R\gg a$ guarantees the potential is smooth on the lattice scale, making intervalley scattering exponentially suppressed and permitting an independent treatment of the two valleys \cite{CastroNetoRMP2009,PeresRMP2010}. The ensemble is characterized by the impurity areal density $n_{imp}$. We work in the dilute limit, $n_{imp} R^2 \ll 1$, which ensures the average separation between impurities is much larger than their spatial extent. Within this framework, the
transport properties are derived from the scattering transition rates induced by this perturbation \cite{NovikovPRB2007,FerreiraPRB2011}. A schematic of randomly distributed soft-sphere potentials on a graphene sheet is shown in Fig.~\ref{fig:Model}.

\section{Scattering through Partial-waves} \label{phase_shifts}
In this section we analyze the scattering of Dirac electrons in graphene from soft sphere potentials by mean of the partial wave formalism. We begin by explicitly solving the corresponding time-independent Dirac equation in both inner (soft sphere potential) and outer (free Graphene states) regions. By enforcing continuity, partial phase shifts are obtained analytically, enabling further numerical analysis of the total phase shift. This quantity will be used in Sec. \ref{transport_section} to obtain an expression for the transport scattering time.

The soft sphere potential describes a finite range perturbation which remains constant within its range $R$, while vanishes for $r>R$, i.e., $V(\bm{r}) = V_0 \big(\Theta(r) - \Theta(r-R)\big)$. The Dirac equation must be solved separately in the inner ($r<R$) and outer ($r>R$) regions. However, since $V(\bm{r})$ is a constant potential, the structure of the solutions is formally identical in both regions. In polar coordinates, Dirac equation yields the coupled equations
\begin{align}
    -i e^{-i \theta} \parenth{ \partd{}{r} - \frac{1}{r} \partd{}{\theta} } \Psi_{2\chi \xi} - k_{\xi} \Psi_{1\chi \xi} = 0,
    \label{Dirac_Eq-Coupled1} \\[6pt]
    -i e^{i \theta} \parenth{ \partd{}{r} + \frac{1}{r} \partd{}{\theta} } \Psi_{1\chi \xi} - k_{\xi} \Psi_{2\chi \xi} = 0,
    \label{Dirac_Eq-Coupled}
\end{align}
where $\xi$ denotes the inner (in) and outer (out) regions. From Graphene's dispersion relation (Eq. \eqref{eq:Disp_Rel}), one finds $k_{\text{out}} = s k$ and $k_{\text{in}} = sk - V_0 / \chi \hbar v_F$. Conservation of the total angular momentum allows the solution to be expressed as a superposition of eigenstates with definite $m_j$, where each partial wave takes the form
\begin{equation}
    \Phi_{\chi\xi}^{m_j} = \begin{pmatrix}
        f^{\chi \xi}_{m_{j}^{-}}(r) e^{i(m_{j}^{-})\theta} \\[6pt]
        i s g^{\chi\xi}_{m_{j}^{+}}(r) e^{i(m_{j}^{+})\theta}
    \end{pmatrix},
\end{equation}
where $m_{j}^{\pm}=m_{j} \pm 1/2$. Substituting the expansion into Eqs. \eqref{Dirac_Eq-Coupled1} and \eqref{Dirac_Eq-Coupled} yields an uncoupled equation for the upper spinor component
\begin{equation}
    \!\!\!\! \parenth{-\frac{1}{r}\partial_r(r\partial_r) + \frac{(m_j - 1/2)^2}{r^2}}  f^{\chi\xi}_{m_{j}^{-}}  = k_{\xi}^2  f^{\chi\xi}_{m_{j}^{-}}.
\end{equation}
The corresponding solution, when substituted back into Eq. \eqref{Dirac_Eq-Coupled}, determines the complete spinor. For the outer region, the solution can be written as
\begin{equation}
    \Phi_{\text{out}}^{m_j} =\begin{pmatrix}
            \left[ c _{1} \, J _{m_{j}^{-}} (kr) + c _{2} \, Y _{m_{j}^{-}} (kr)  \right]\! e ^{i (m_{j}^{-}) \theta } \\[6pt] \! i s  \!\left[ c _{1} \, J _{m_{j}^{+}} (kr) + c _{2} \, Y _{m_{j}^{+}} (kr) \right] \!  e ^{i (m_{j}^{+}) \theta }
        \end{pmatrix} ,
\end{equation}
where $J_\nu$, $Y_\nu$ denote the Bessel functions of the first and second kind, respectively. The inner-region solution is obtained in an analogous manner. However, since  $Y_{\nu}$ diverges at the origin, only the first kind Bessel function contributes. The phase shifts are determined by enforcing continuity at $r=R$. A direct calculation yields
\begin{equation}
    \tan \delta_{m_j} = \frac{ J _{m_{j}^{-}} (kR) \, J _{m_{j}^{+}} \parenth{|k_{\text{in}}|R} - J _{m_{j}^{+}} (kR) \, J _{m_{j}^{-}} \parenth{|k_{\text{in}}|R}}{ Y _{m_{j}^{-}} (kR) \, J _{m_{j}^{+}} \parenth{|k_{\text{in}}|R} - Y _{m_{j}^{+}} (kR) \, J _{m_{j}^{-}} \parenth{|k_{\text{in}}|R} }.
    \label{eq:Phase_shiftC}
\end{equation}
Since $|k_{\text{in}}|$ depends only on the product $s\chi$ fixing the electronic band also fixes it, implying that the scattering is independent of the valley index. Once the phase shifts are known, the transition matrix $T$ can be expressed in terms of them. Its squared modulus is given by \cite{S_Ramezani_2011,CBM_R2025}
    \begin{equation}
    |T_{\bm{k'}\bm{k}}|^2 = \frac{4\hbar^2 v_F^2}{k^2}\!\! \sum_{m_j,m_{j}^{\prime}}e^{i(m_j-m_{j}^{\prime})\theta}e^{i(\delta_{m_j}-\delta_{m_{j}^{\prime}})}\!\sin\delta_{m_j}\sin\delta_{m_{j}^{\prime}}.\label{eq:T_phase_shifts}
    \end{equation}
Equation \eqref{eq:T_phase_shifts} provides an exact solution to the Lippmann–Schwinger equation, allowing the quantum transition rate to be evaluated without approximation. In contrast, a widely used and simpler approach is to employ the Born approximation, in which the $T$ matrix is expressed directly in terms of the Fourier transform of the scattering potential \cite{PhysRevB.91.035202,PhysRevLett.96.256602,PhysRevB.84.235126}. Such framework yields analytical expressions for the scattering amplitude and transport coefficients, and its computational simplicity makes it attractive for studying disorder-averaged properties. However, the partial-waves based approach employed here remains essential for a quantitatively reliable description of scattering, being accurate even for strong potentials or low-energy regimes, thus providing a physically realistic and quantitatively accurate framework for understanding how extended impurities reshape charge and energy currents in disordered Dirac materials \cite{PeresRMP2010,DasSarmaRMP2011}.

\section{Semiclassical transport analysis} \label{transport_section}
Transport properties are analyzed within the semiclassical Boltzmann formalism under the relaxation-time approximation. In this framework, the transport scattering time is derived from Fermi's golden rule, which quantifies scattering probabilities between elastic states labeled by band index $s=\pm1$ and valley index $\chi=\pm1$, with linear dispersion $\mathscr{E}_{s\chi}(k)=s\,\chi\,\hbar v_F k$. The dilute impurity limit enables averaging over disorder configurations, introducing a factor of the impurity density $n_{\mathrm{imp}}$, reflecting the number of independent scattering centers per unit area. The resulting transition rate between states $\bm{k}$ and $\bm{k}^{\prime}$ is given by \cite{sakurai1994modern}
\begin{align}
    W_{\bm{k'}\bm{k}}= \frac{2 \pi}{\hbar} n_{\mathrm{imp}} |T_{\bm{k'}\bm{k}}|^2 \delta\parenth{\mathscr{E}_{s\chi}(\bm{k})-\mathscr{E}_{s\chi}(\bmp{k})}.
    \label{eq:TrRate}
\end{align}
The transport scattering time is obtained by weighting these processes with the angular factor $(1-\cos(\theta))$, which quantifies momentum relaxation efficiency
\begin{align}
    \frac{1}{\tau_{\mathrm{tr}}\big(\mathscr{E}_{s\chi}(\bm{k})\big)} =
    \int \frac{d^{2}k'}{(2\pi)^{2}}\; (1-\cos\theta)\;
    W_{\bm{k'}\bm{k}}.
\end{align}
Substitution of Eqs. \eqref{eq:TrRate} and \eqref{eq:T_phase_shifts} enables direct evaluation of the expression above. Integration on the radial variable is carried out using the two-dimensional Dirac density of states $\nu_{1}\big(\mathscr{E}_{s\chi}\big)=k/(2\pi \hbar v_F)$, while angular integration leads to selection of adjacent angular momentum channels only. The final expression is
\begin{align}
    \frac{1}{\tau_{\mathrm{tr}}\big( \bm{k} \big)} =
     \frac{2 n_{\mathrm{imp}} v_F}{k} \!\!
    \sum_{m=-\infty}^{\infty}    \sin^{2}\!\Big[\delta_{m+1}(\bm{k})-\delta_{m}(\bm{k})\Big].
    \label{eq:tautr_phases}
\end{align}
Equation~\eqref{eq:tautr_phases} reveals a distinctive property of Dirac carriers: momentum relaxation depends on differences between adjacent phase shifts. This behavior directly manifests pseudospin-momentum locking, which suppresses exact backscattering from smooth scalar potentials \cite{NovikovPRB2007}.

For carriers labeled by spin $\sigma$, band index $s$, and valley $\chi$, the distribution function satisfies \cite{ziman}
\begin{equation}
    \dot{\bm{x}}\cdot\nabla_{\bm{x}} f_{\sigma s\chi}
    + \dot{\bm{k}}\cdot\nabla_{\bm{k}} f_{\sigma s\chi}
    = -\frac{g_{\sigma s\chi}}{\tau_{\mathrm{tr}}(\bm{k})},
    \label{BE_BoltzmannEq}
\end{equation}
where $g_{\sigma s\chi} = f_{\sigma s\chi} - f^0_{\sigma s\chi}$ measures the deviation from the local Fermi–Dirac equilibrium.
Neglecting Berry curvature, the semiclassical dynamics take the standard form
\begin{equation}
    \dot{\bm{x}} = \frac{1}{\hbar}\nabla_{\bm{k}} \mathscr{E}_{s\chi},
    \quad
    \dot{\bm{k}} = \frac{q}{\hbar}(\bm{E} + \dot{\bm{x}}\times\bm{B}).
    \label{BE_SemClasDyn}
\end{equation}
For $\bm{B}=0$, substituting eq. \eqref{BE_SemClasDyn} into eq. \eqref{BE_BoltzmannEq}, and under weak perturbations, the linear order deviation becomes
\begin{align}
    g_{\sigma s\chi} =& \tau_{\mathrm{tr}}(\bm{k})
    \left( -\frac{\partial f^0}{\partial \mathscr{E}} \right)
    \bm{v}_{s\chi}(\bm{k}) \notag\\
    &\times \left[ q\!\left(\bm{E} - \frac{\nabla\mu}{q}\right)
    + \frac{\mathscr{E}_{s\chi} - \mu}{T}(-\nabla T) \right],
    \label{BE_LODist}
\end{align}
with $\bm{v}_{s\chi} = \hbar^{-1} \nabla_{\bm{k}} \mathscr{E}_{s\chi}$. The first term inside brackets corresponds to the electrochemical driving field, and the second to the thermal driving force. Following eq. \eqref{BE_LODist}, the carrier driven electric ($\bm{J}^1$) and heat ($\bm{J}^2$) currents can be written as
\begin{align}
    \bm{J}^1 &= \frac{q^2}{T} \bm{L}^{(11)}\!\left(\bm{E} - \frac{\nabla\mu}{q}\right)
    + \frac{q}{T^2} \bm{L}^{(12)}(-\nabla T),
    \label{eq:Elec_Current}\\
    \bm{J}^2 &= \frac{q}{T} \bm{L}^{(21)}\!\left(\bm{E} - \frac{\nabla\mu}{q}\right)
    + \frac{1}{T^2} \bm{L}^{(22)}(-\nabla T),\label{eq:Therm_Current}
\end{align}
where $\bm{L}^{(ij)}$ denotes the Onsager coefficients, defined by the expression
\begin{align}
    \bm{L}^{(ij)} =&\; T\!\sum_{\sigma,s,\chi} \!\int \frac{d^2k}{(2\pi)^2}
    \tau_{\mathrm{tr}}(\bm{k})
    [\mathscr{E}_{s\chi} - \mu]^{i+j-2} \notag\\
    &\times\!\left(-\frac{\partial f^0}{\partial \mathscr{E}}\right)
    \bm{v}_{s\chi} \otimes \bm{v}_{s\chi}.
\end{align}

For graphene, $\bm{v}_{s\chi} = s\chi v_F \hat{k}$ makes $\bm{L}^{(ij)}$ diagonal. Spin and valley degeneracies contribute a factor of $4$, while Fermi-surface selection restricts the band sum. In the $T\!\to\!0$ limit, Fermi-Dirac distribution leads to $-\partial f^0/\partial\mathscr{E} \to \delta(\mathscr{E}-\mathscr{E}_F)$, enabling closed-form evaluation. Within the low temperature regime ($k_B T \ll \mathscr{E}_F$), the Sommerfeld expansion \cite{ziman_principles} leads to 
\begin{equation}
    L^{(11)}_{\alpha\beta} = \frac{T\mathscr{E}_F}{\pi\hbar^2}\delta_{\alpha\beta}
    \!\left[ \tau_{\mathrm{tr}} + \frac{(\pi k_B T)^2}{6\mathscr{E}_F}
    ( 2\tau_{\mathrm{tr}}' + \mathscr{E}_F \tau_{\mathrm{tr}}'' ) \right],\label{eq:L11}
    \end{equation}
    \begin{equation}
    L^{(12)}_{\alpha\beta} = \frac{\pi k_B^2 T^3}{3\hbar^2} \delta_{\alpha\beta}
    \!\left( \tau_{\mathrm{tr}} + \mathscr{E}_F\tau_{\mathrm{tr}}' \right),\label{eq:L12}
    \end{equation}
    \begin{equation}
    L^{(22)}_{\alpha\beta} = \frac{\pi k_B^2 T^3}{3\hbar^2} \delta_{\alpha\beta}
    \mathscr{E}_F \tau_{\mathrm{tr}}\label{eq:L22},
\end{equation}
where all $\tau_{\mathrm{tr}}$ and derivatives are evaluated at $\mathscr{E}_F$. The main transport-related quantities, within the low-temperature regime, are derived from the above expressions for the Onsager coefficients.

\paragraph{Electrical conductivity}. The DC electrical conductivity, $\bm{\sigma}^{\mathrm{DC}}$, describes the electrical response under isothermal conditions ($\nabla T = 0$). It is obtained by substituting Eq.~\eqref{eq:L11} into Eq.~\eqref{eq:Elec_Current}, which yields
    \begin{equation}
    \sigma^{\mathrm{DC}}_{\alpha\beta} = \frac{q^2 \mathscr{E}_F}{\pi\hbar^2}\delta_{\alpha\beta}
    \!\left[ \tau_{\mathrm{tr}} + \frac{(\pi k_B T)^2}{6\mathscr{E}_F}
    ( 2\tau_{\mathrm{tr}}' + \mathscr{E}_F \tau_{\mathrm{tr}}'' ) \right],
    \label{eq:cond_DC_final}
    \end{equation}
In the zero-temperature limit, this expression simplifies to the residual electrical conductivity
\begin{equation}
    \sigma_{\alpha\beta}(0) = 2\frac{q^2}{h}\,
    \frac{\mathscr{E}_F}{\hbar}\tau_{\mathrm{tr}} \, \delta_{\alpha\beta}.
    \label{eq:cond_DC_zero}
\end{equation}
\paragraph{Electronic thermal conductivity.} The electronic thermal conductivity, $\bm{\kappa}^{\mathrm{el}}$, is the thermal response under the condition of zero electrical current ($\bm{J}^{1} = 0$). This condition is used to express the electric field in terms of the temperature gradient, which, upon substitution into Eq.~\eqref{eq:Therm_Current}, yields the general expression
\begin{equation}
    \bm{\kappa}^{\mathrm{el}} = \frac{1}{T^2}\left[
    \bm{L}^{(22)} - \bm{L}^{(12)}(\bm{L}^{(11)})^{-1}\bm{L}^{(21)}\right].
\end{equation}
To leading order in temperature, the electronic thermal conductivity is given by
\begin{equation}
    \kappa^{\mathrm{el}}_{\alpha\beta}
    \approx \frac{\pi k_B^2 T}{3\hbar^2} \, \mathscr{E}_F \tau_{\mathrm{tr}} \, \delta_{\alpha\beta}. \label{eq:cond_term_final}
\end{equation}
\paragraph{Lorenz coefficient.} A comparison between Eqs.~\eqref{eq:cond_DC_zero} and~\eqref{eq:cond_term_final} confirms the validity of the Wiedemann-Franz law in the $T \to 0$ limit. The Lorenz number, $L_{\alpha\beta} = \kappa^{\mathrm{el}}_{\alpha\beta} / (T \sigma^{\mathrm{DC}}_{\alpha\beta})$, quantifies deviations from this ideal metallic behavior at finite temperatures. Substituting the low-temperature expressions for $\sigma^{\mathrm{DC}}_{\alpha\beta}$ and $\kappa^{\mathrm{el}}_{\alpha\beta}$ gives
\begin{align}
    L_{\alpha\beta} = \frac{ L_0 \, \delta_{\alpha\beta}}{1 + \frac{(\pi k_B T)^2}{6\mathscr{E}_F\tau_{\mathrm{tr}}}
    ( 2\tau_{\mathrm{tr}}' + \mathscr{E}_F \tau_{\mathrm{tr}}'' )}, \label{eq:LNumber}
\end{align}
where $L_0 = (\pi^2/3)(k_{_B}/e)^2$ is the Sommerfeld value.
\paragraph{Seebeck coefficient.} The electrical voltage generated in response to a thermal gradient, under open-circuit conditions ($\bm{J}^{1}=0$), is quantified by the Seebeck coefficient (or thermopower), $\bm{S}$. It can be calculated by setting $\bm{J}^{1}=0$ in Eq. \eqref{eq:Elec_Current}, yielding the relation
\begin{align}
    \bm{S} = \frac{1}{qT} \parenthc{\bm{L}_{(11)}}^{-1} \bm{L}_{(12)}.
\end{align}
Using the low-temperature expansions for the Onsager coefficients (Eqs.~\eqref{eq:L11} and~\eqref{eq:L12}), we obtain:
\begin{align}
    S_{\alpha\beta} = \frac{\pi^2 k_B^2 T}{3 q \mathscr{E}_F}  \frac{\parenth{\tau_{\mathrm{tr}} + \mathscr{E}_F\tau_{\mathrm{tr}}'} \delta_{\alpha\beta}}{\tau_{\mathrm{tr}} + \frac{(\pi k_B T)^2}{6\mathscr{E}_F}
    ( 2\tau_{\mathrm{tr}}' + \mathscr{E}_F \tau_{\mathrm{tr}}'' )}
    \label{eq:Seebeck}
\end{align}
\paragraph{Figure of merit.} The overall thermoelectric efficiency of a material is quantified by the dimensionless figure of merit, $ZT$, defined as:
\begin{align}
    ZT = \frac{S^2\sigma T}{\kappa}
\end{align}
This dimensionless quantity directly correlates the material's electrical conductivity ($\sigma$), its thermal conductivity ($\kappa$), and the magnitude of the thermopower $S$. 
Although $\kappa$ represents the total thermal conductivity, which includes both electronic and phononic contributions, in this work we evaluate $ZT$ considering $\kappa \approx \kappa^{\mathrm{el}}$ to provide an upper bound on the electronic efficiency and to clarify how smooth disorder specifically governs the energy flow of Dirac fermions.

\section{Results and Discussion} \label{results_discussion_section}
In this section, we evaluate numerically the DC and thermal conductivities, as well as the thermoelectric quantities.  Soft sphere potentials emerge as a simple model in the study of localized short-range impurities \cite{Dean2010}, though supercritically charged vacancies are also described in a similar form \cite{Mao2016}. Following \cite{Dean2010}, the diameter of the spheres is on the order 3 nm, and are randomly distributed with a density of $n_{\mathrm{imp}}\sim 1 \times10^{12}$ cm$^{-2}$. The Fermi velocity is set to $v_F \sim 10^{6}$ m/s, according with previously reported values for graphene \cite{DasSarma2011ElectronicTransport}, also, the charge carriers considered are electrons. Then, we select the conduction band ($s\chi = +1$) and set the electronic charge as $q = -e = -1.602\times10^{-19}$ C. For graphene, standard values of the Fermi energy are of the order $\mathscr{E}_F\sim 200$ meV, corresponding to a density of carriers $n_c\sim 10^{12}$ cm$^{-2}$ \cite{DasSarma2011ElectronicTransport}. This choice of parameters agrees with the commonly accepted range of validity for the linear model in graphene, which remains accurate for energies as high as $1\,\text{eV}$. The minimum conductivity measured in graphene, taking into account the spin and valley symmetry, is $\sigma_{xx}= 4 e^2/h\approx 115\, \mu$S \cite{Tan2007Measurement,Kim2016ValleySymmetry}.

\begin{figure}[b]
    \centering
    \begin{subfigure}{0.45\textwidth}
         \centering
         \includegraphics[width=\linewidth]{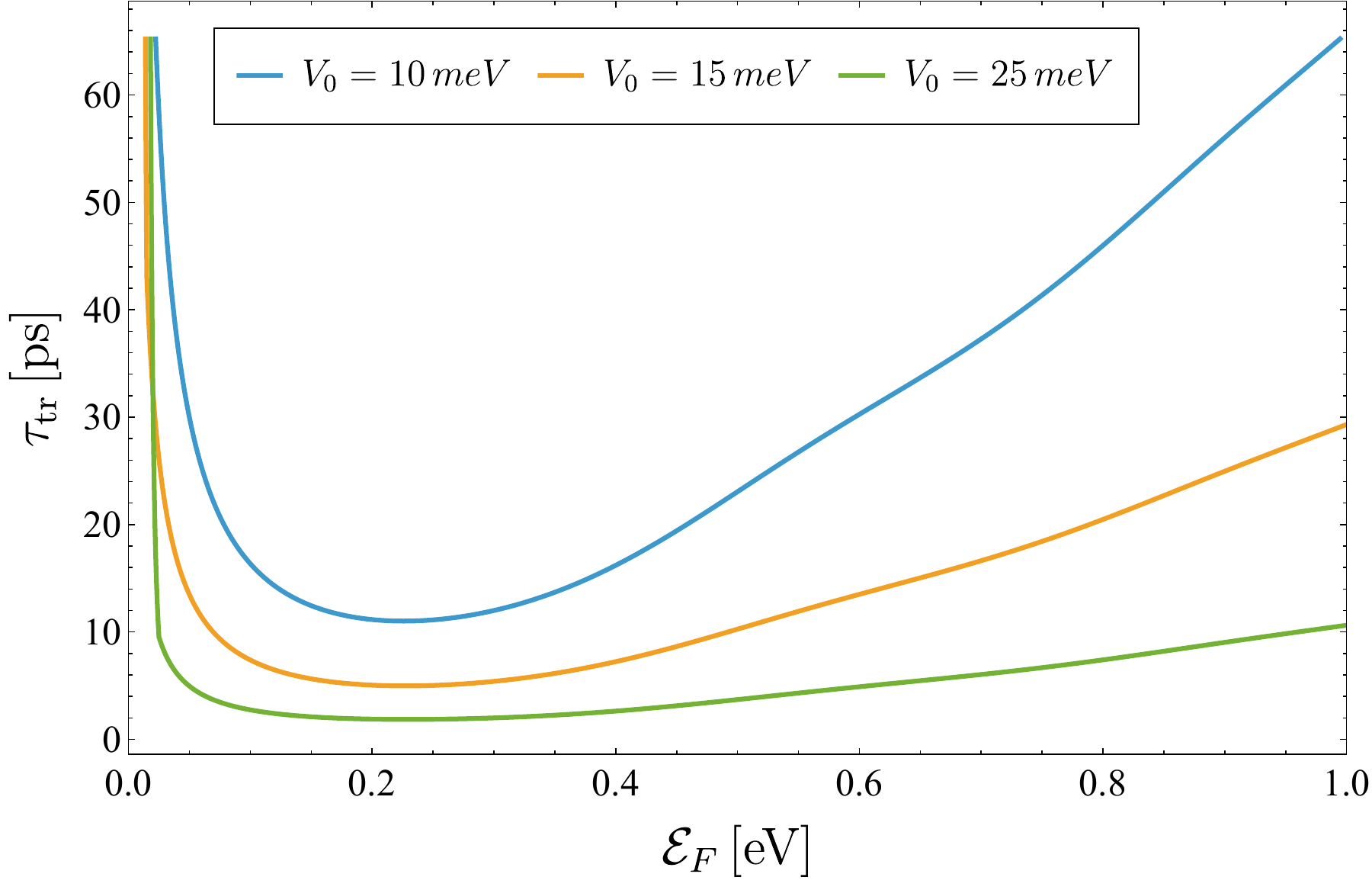}
         \subcaption{}\label{fig:tau_pos}
     \end{subfigure}
     \begin{subfigure}{0.45\textwidth}
         \centering
         \includegraphics[width=\linewidth]{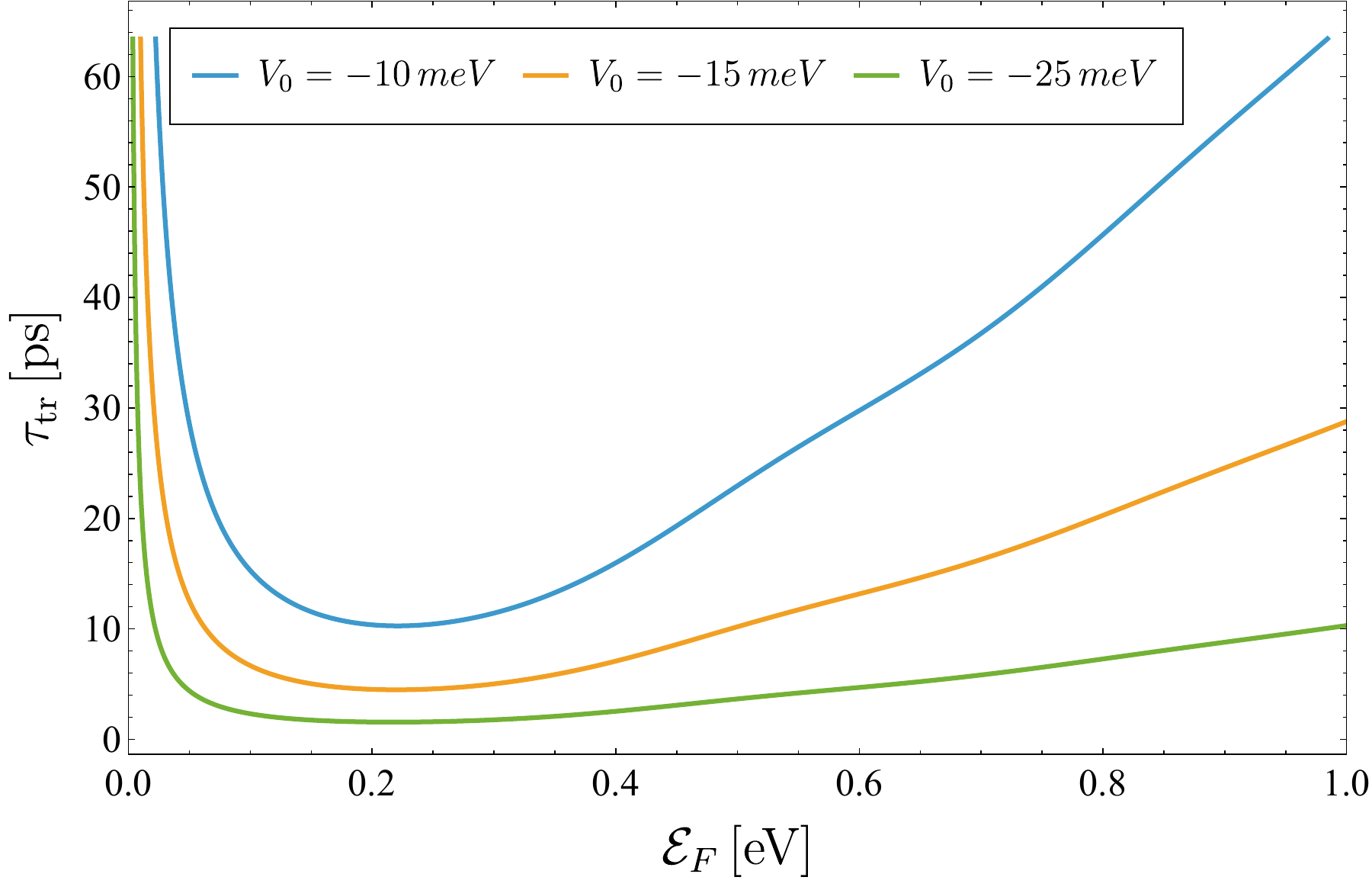}
         \subcaption{}\label{fig:tau_neg}
     \end{subfigure}
    \caption{Transport relaxation time (in picoseconds) as a function of the Fermi energy, calculated using Eq.~\eqref{eq:tautr_phases} for a random distribution of 3 nm-radius soft spheres with a concentration of $n_{\mathrm{imp}} = 1 \times 10^{12}$ cm$^{-2}$, and for various potentials. The subfigure (a) corresponds to positive soft sphere potentials, whereas (b) corresponds to negative ones.} 
 \label{fig:tau_vs_EF}
\end{figure}
\paragraph{The transport relaxation time.} Fig.~\ref{fig:tau_vs_EF} shows the behavior of the relaxation time as a function of the Fermi energy, computed using Eq.~\eqref{eq:tautr_phases}, for several negative and positive values of the soft sphere potential. The figure reveals a monotonic increase in the relaxation time with the Fermi energy. This behavior contrasts with that reported in ref. \cite{CBM_R2025}, where a random distribution of Dirac oscillators is shown to induce pronounced minima at specific Fermi energies. Those minima were attributed to resonant scattering via quasi-bound states associated with Landau levels. The absence of such features in the case of soft-sphere scatterers suggests that this type of potential lacks the capacity to form analogous quasi-bound states, leading to the observed featureless, monotonic dependence. 

Furthermore, the relaxation time is observed to decrease as the strength of the soft-sphere potential is increased. This trend is consistent with the expectation that a stronger scattering potential enhances the scattering rate by deflecting charge carriers more effectively. Crucially, even at higher potentials, no minima emerge in the spectrum. This absence further corroborates the hypothesis that the soft-sphere potential, regardless of its strength, is fundamentally incapable of supporting the formation of quasi-bound states, which would be necessary for resonant scattering to occur.

\begin{figure}[!ht]
\centering
    \begin{subfigure}[]{0.45\textwidth}
         \centering
         \includegraphics[width=\linewidth]{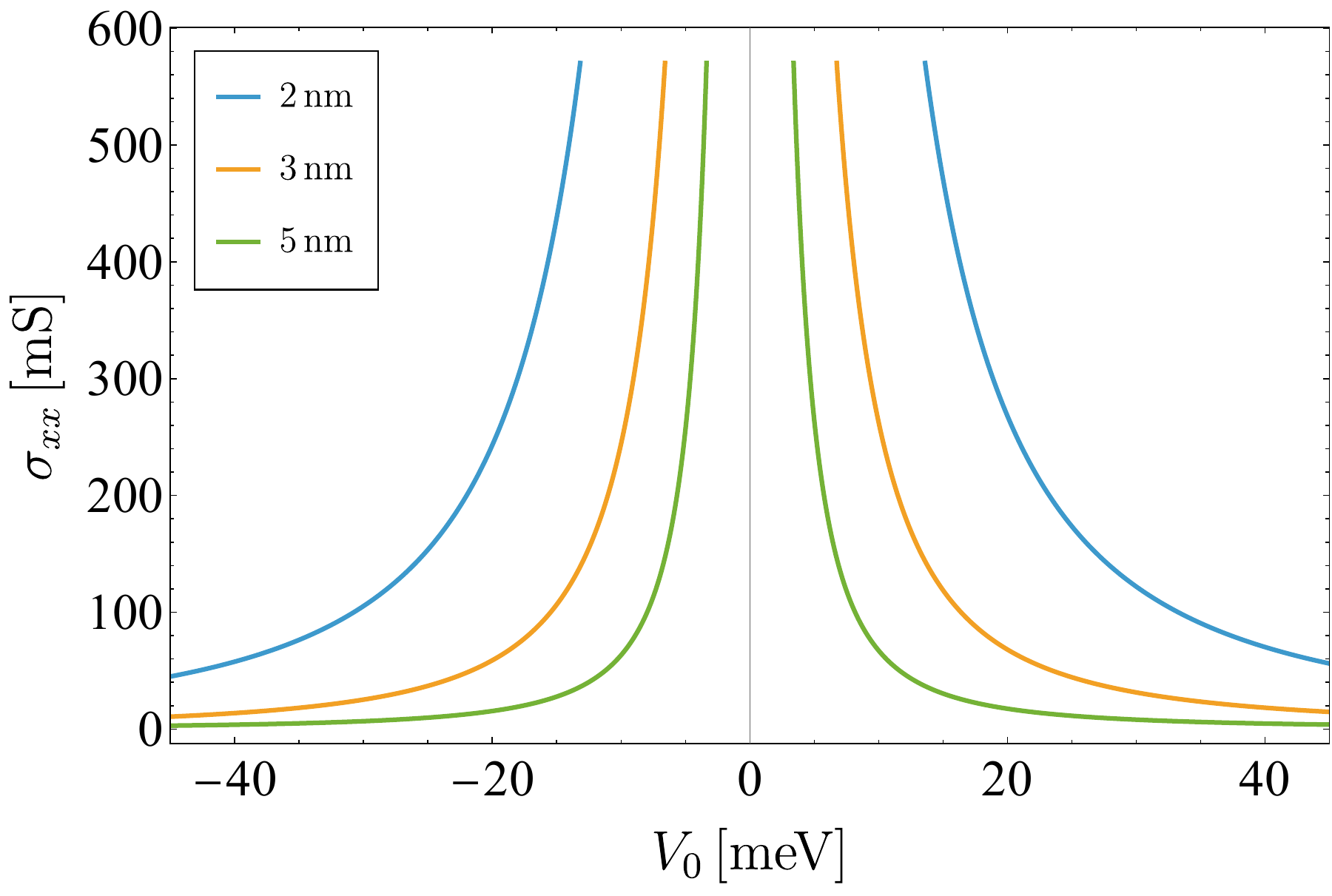}
         \subcaption{}\label{fig:cond_vs_R}
     \end{subfigure}
     \begin{subfigure}[]{0.45\textwidth}
         \centering
         \includegraphics[width=\linewidth]{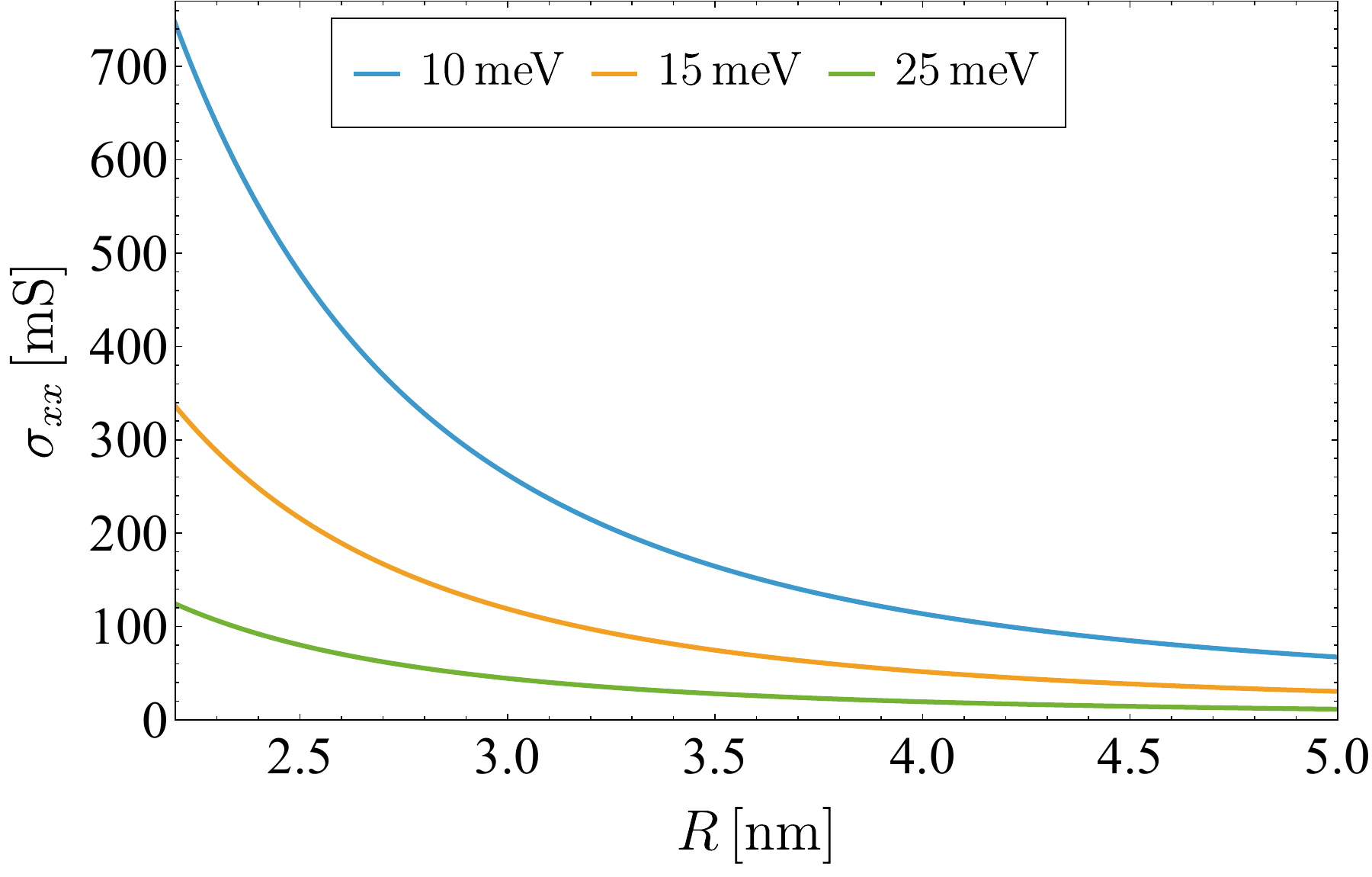}
         \subcaption{}\label{fig:cond_vs_B}
     \end{subfigure}
    \caption{DC conductivity at zero temperature versus (a) the radii of the spheres  and (b) the soft-sphere potential. In both cases, the Fermi energy is $\mathscr{E}_F = 200$ meV.} 
    \label{fig:cond_vs_R_B}
\end{figure}
\paragraph{Electrical and thermal conductivities.} We now turn to the analysis of the DC electrical conductivity, $\sigma_{xx}$, and the electronic thermal conductivity, $\kappa_{xx}$. Given that both the electrical and thermal conductivity tensors are diagonal and isotropic, we restrict our discussion to these longitudinal components. As shown in Eq.~\eqref{eq:tautr_phases}, the transport relaxation time decreases inversely with the nanobubble density. Accordingly, both the electrical and thermal conductivities can be modulated through the concentration of nanobubbles. A tenfold increase in their density results in an approximately one-order-of-magnitude reduction of both conductivities. Figure~\ref{fig:cond_vs_R} presents the residual DC conductivity as a function of the sphere radius. The conductivity exhibits an overall decrease with increasing radius. Since $\sigma_{xx}$ is directly proportional to the transport relaxation time, this monotonic behavior is consistent with the absence of resonant features.

Figure~\ref{fig:cond_vs_B} displays the conductivity as a function of the soft-sphere potential strength. At first glance, the conductivity appears to be symmetric with respect to the sign of $V_0$. However, Eq. \eqref{eq:Phase_shiftC} reveals that this symmetry is not exact, as the wavevector inside the potential region, $|k_\text{in}|$, itself depends on the sign of $V_0$. The apparent symmetry observed in the plot is a consequence of the strikingly similar behavior of the transport relaxation time for both positive and negative potentials of equal magnitude, as previously shown in Fig. \ref{fig:tau_vs_EF}. The results further demonstrate that the conductivity is suppressed as the magnitude of the potential $|V_0|$ is increased. Notably, this reduction becomes less pronounced for larger values of $|V_0|$, indicating a saturation-like effect in the scattering strength.

\begin{figure}[!ht]
    \centering
    \begin{subfigure}[]{0.45\textwidth}
         \centering
         \includegraphics[width=\linewidth]{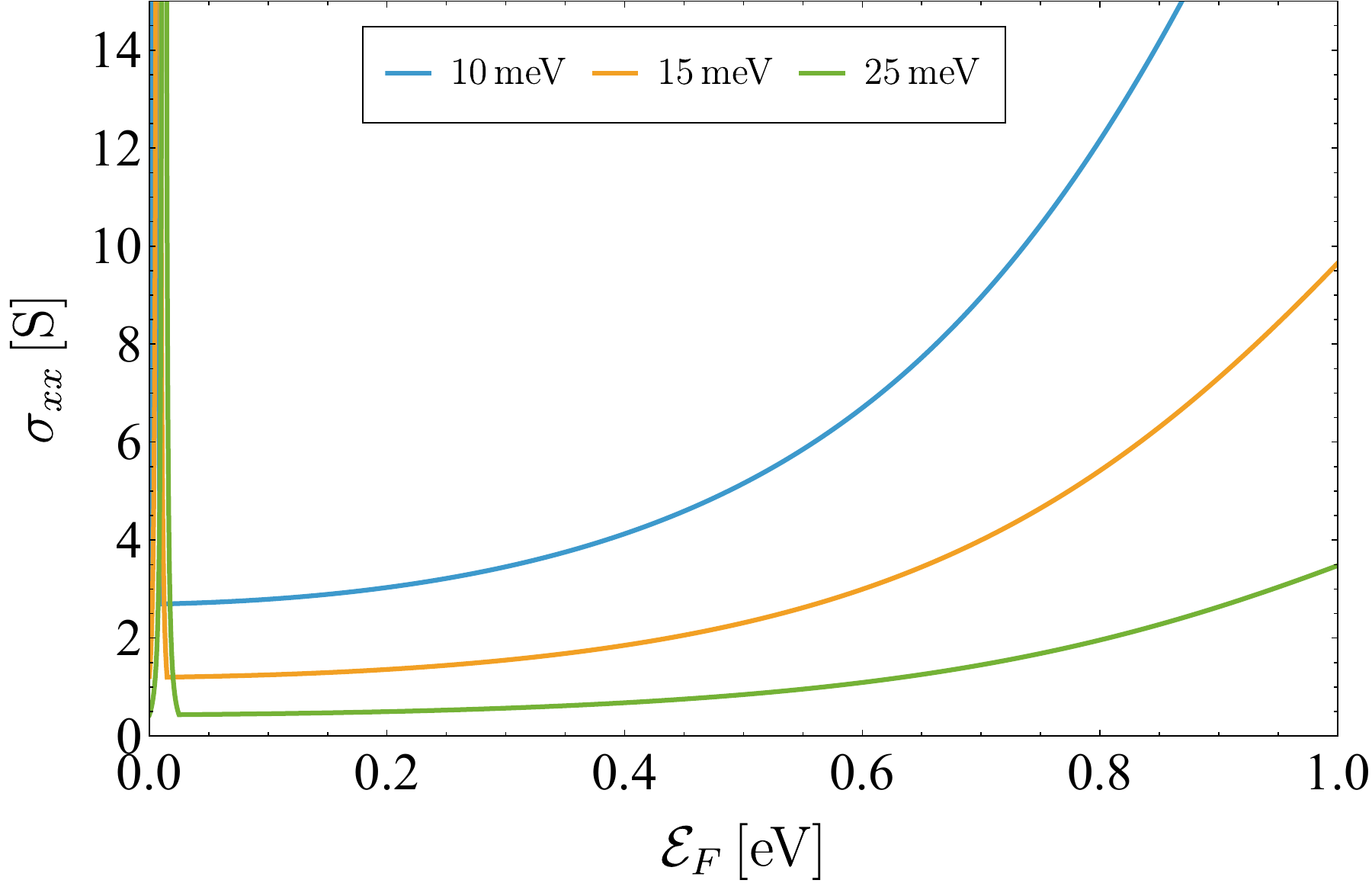}
         \subcaption{}\label{fig:cond_vs_EF_R2}
     \end{subfigure}
     \begin{subfigure}[]{0.45\textwidth}
         \centering
         \includegraphics[width=\linewidth]{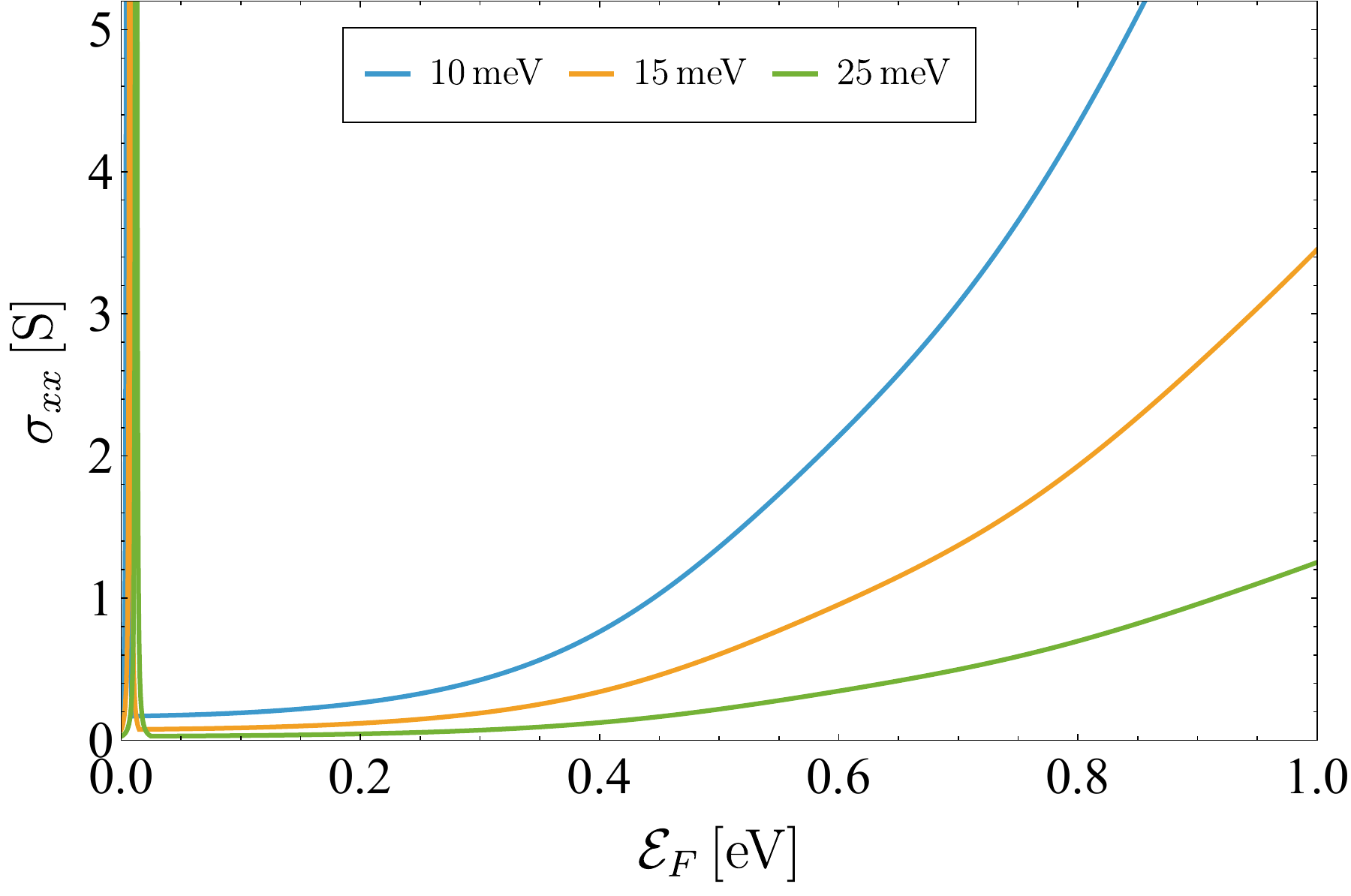}
         \subcaption{}\label{fig:cond_vs_EF_R4}
     \end{subfigure}
      \begin{subfigure}[]{0.45\textwidth}
         \centering
         \includegraphics[width=\linewidth]{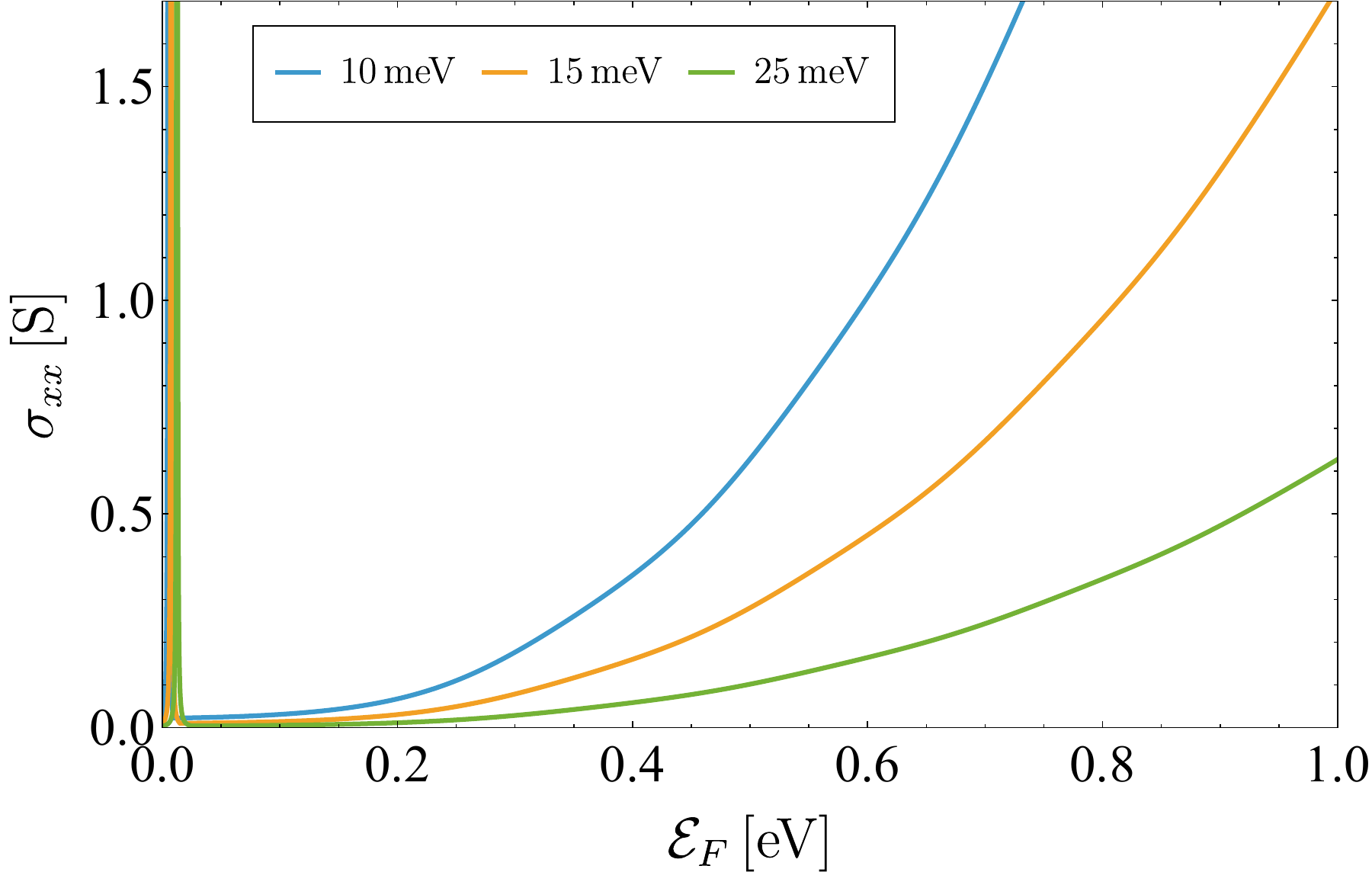}
         \subcaption{}\label{fig:cond_vs_EF_R6}
     \end{subfigure}
    \caption{DC conductivity versus the Fermi energy at zero temperature for a distribution of: (a) 1.5 nm-radius spheres, (b) 3 nm-radius spheres and (c) 5 nm-radius spheres.} 
    \label{fig:cond_vs_EF}
\end{figure}
Figure~\ref{fig:cond_vs_EF} presents the residual electrical conductivity as a function of the Fermi energy, computed from Eq.~\eqref{eq:cond_DC_zero}, for three representative values of the soft-sphere radius. A key feature is the divergent trend in conductivity observed at low Fermi energies for smaller sphere radii and weaker potentials. This behavior can be explained by the effective transparency of the scattering centers to charge carriers in this regime, a direct analogue of the Ramsauer-Townsend effect observed in the low-energy electron scattering from noble gas atoms \cite{sakurai1994modern}. It is crucial to note that this transparency arises from interference, not from resonant trapping, which would instead produce a maximum in the scattering rate. Notably, this effect is suppressed as either the radius or the strength of the soft-sphere potentials increases, indicating that scattering remains significant even at low energies. Furthermore, and consistent with our analysis of the relaxation time, no sharp resonant minima are present in the conductivity spectra (Figs.~\ref{fig:cond_vs_EF_R4}-\ref{fig:cond_vs_EF_R6}). Furthermore, it is noteworthy that the residual conductivity profile obtained here mirrors the one resulting from a combined model of short-range defects and charged impurities \cite{HwangPRL2007}. This agreement can be interpreted as a validation of the soft-sphere model: it effectively generalizes the short-range description while providing a consistent, simplified approximation to the more complex charged-impurity scenario.

At a qualitative level, our findings are also consistent with transport studies that analyze the impact of different impurity types and disorder profiles on electronic and optical responses in bilayer graphene systems \cite{OptMat2016,EPJB2015} and other graphene based materials \cite{IJMPB2015, OQE2024,ApplPhysA2024a,ApplPhysA2025}. Although bilayer graphene hosts distinct band structures and symmetry properties compared to monolayer graphene, precluding a direct quantitative comparison, these works similarly emphasize the sensitivity of transport coefficients to the spatial extent, screening, and effective range of impurity potentials. In this sense, our results complement these studies by providing a controlled, nonperturbative description of how smooth finite-range disorder governs charge and energy transport in monolayer graphene.

\begin{figure}[!ht]
    \centering
     \begin{subfigure}[]{0.45\textwidth}
         \centering
         \includegraphics[width=\linewidth]{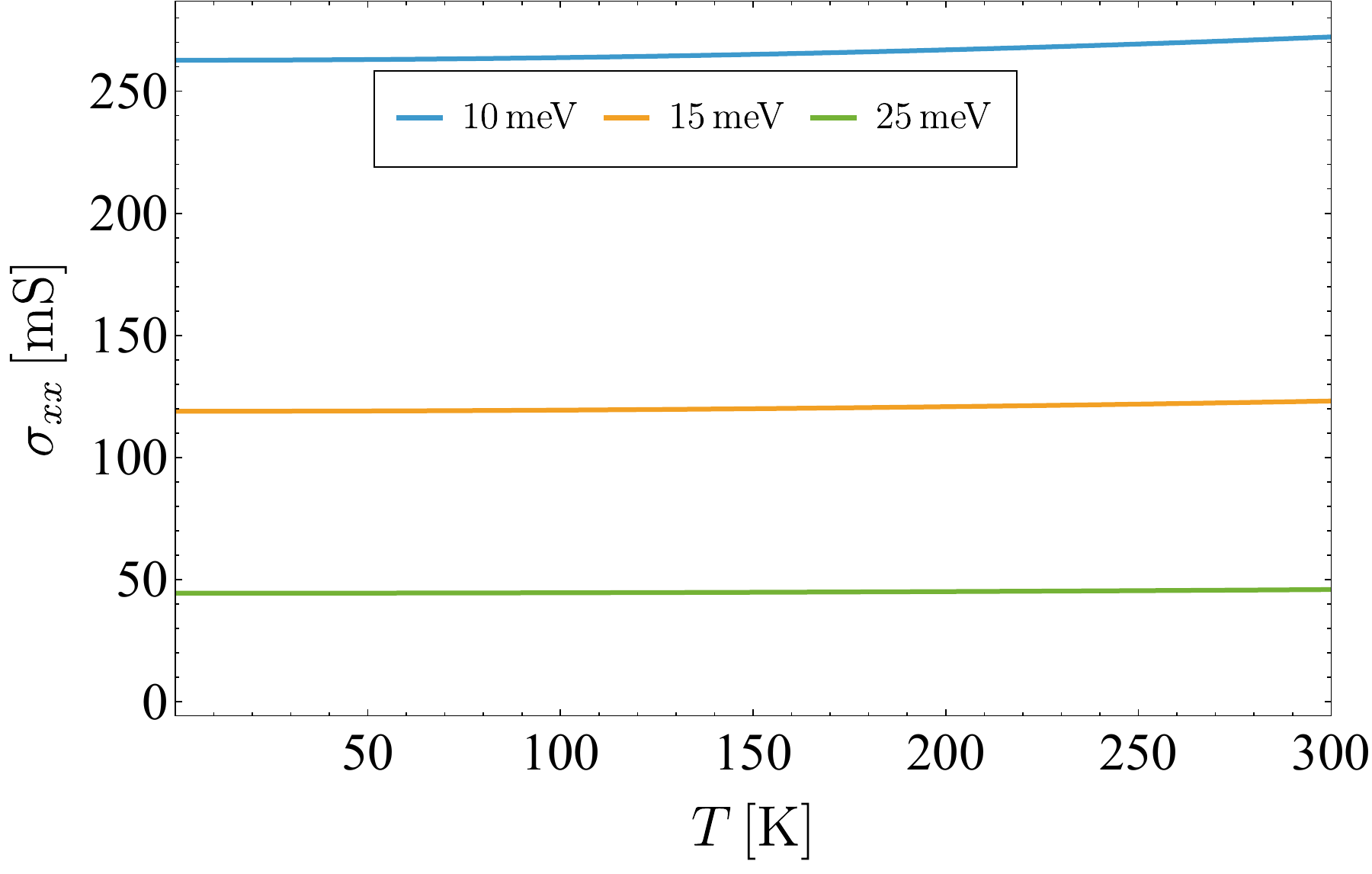}
         \subcaption{}\label{fig:cond_vs_T}
     \end{subfigure}
     \begin{subfigure}[]{0.45\textwidth}
         \centering
         \includegraphics[width=\linewidth]{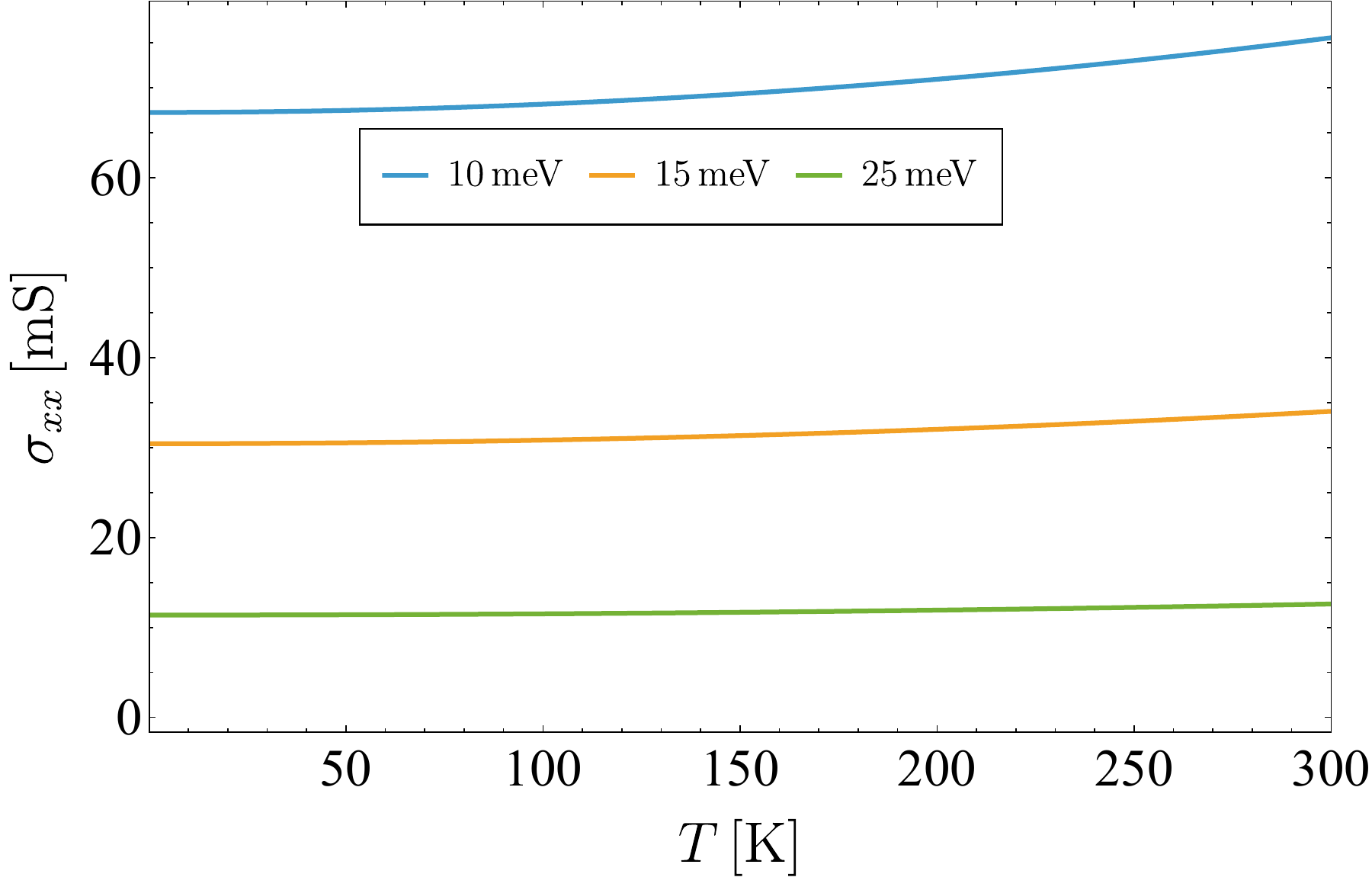}
         \subcaption{}\label{fig:cond_vs_T_EF2}
     \end{subfigure}
    \caption{Temperature dependence of the DC conductivity. The conductivity was calculated at a Fermi energy of $\mathscr{E}_F = 200$ meV, for spheres distributed with  density $n_{\mathrm{imp}} = 1 \times 10^{12}$ cm$^{-2}$ and with: (a) a $3$ nm-radius and (b) a $5$ nm-radius.}
    \label{fig:cond_vs_Temp}
\end{figure}
The temperature dependence of the electrical conductivity, computed from Eq.~\eqref{eq:cond_DC_final}, is presented in Fig.~\ref{fig:cond_vs_Temp}. The structure of the equation indicates that the conductivity is directly influenced by the energy derivative of the transport relaxation time. Given the monotonic increase of $\tau_{tr}$ with energy (Fig.~\ref{fig:tau_vs_EF}), its derivative is positive, thus inherently favoring an increasing or nearly constant conductivity with temperature. The calculated behavior, shown in Figs.~\ref{fig:cond_vs_T}–\ref{fig:cond_vs_T_EF2}, confirms this trend and reveals its parameter dependence. The conductivity is highly sensitive to the strength of the scattering potential ($V_0$), showing a nearly constant profile for large potentials. In contrast, it is shown that the increasing behavior can be tuned by setting large sphere radii. Interestingly, the influence of the Fermi energy ($\mathscr{E}_{F}$) is less pronounced than might be expected. The overall shape of the conductivity vs. temperature curve remains largely unchanged for different $\mathscr{E}_{F}$, primarily shifting upward in magnitude as the Fermi energy is increased. Consequently, the sphere radius acts as the primary parameter modulating the functional form of $\sigma(T)$, while the Fermi energy predominantly sets its overall magnitude near $T=0$ K.

    \begin{figure}[!ht]
    \centering
    \begin{subfigure}[]{0.45\textwidth}
         \centering
         \includegraphics[width=\linewidth]{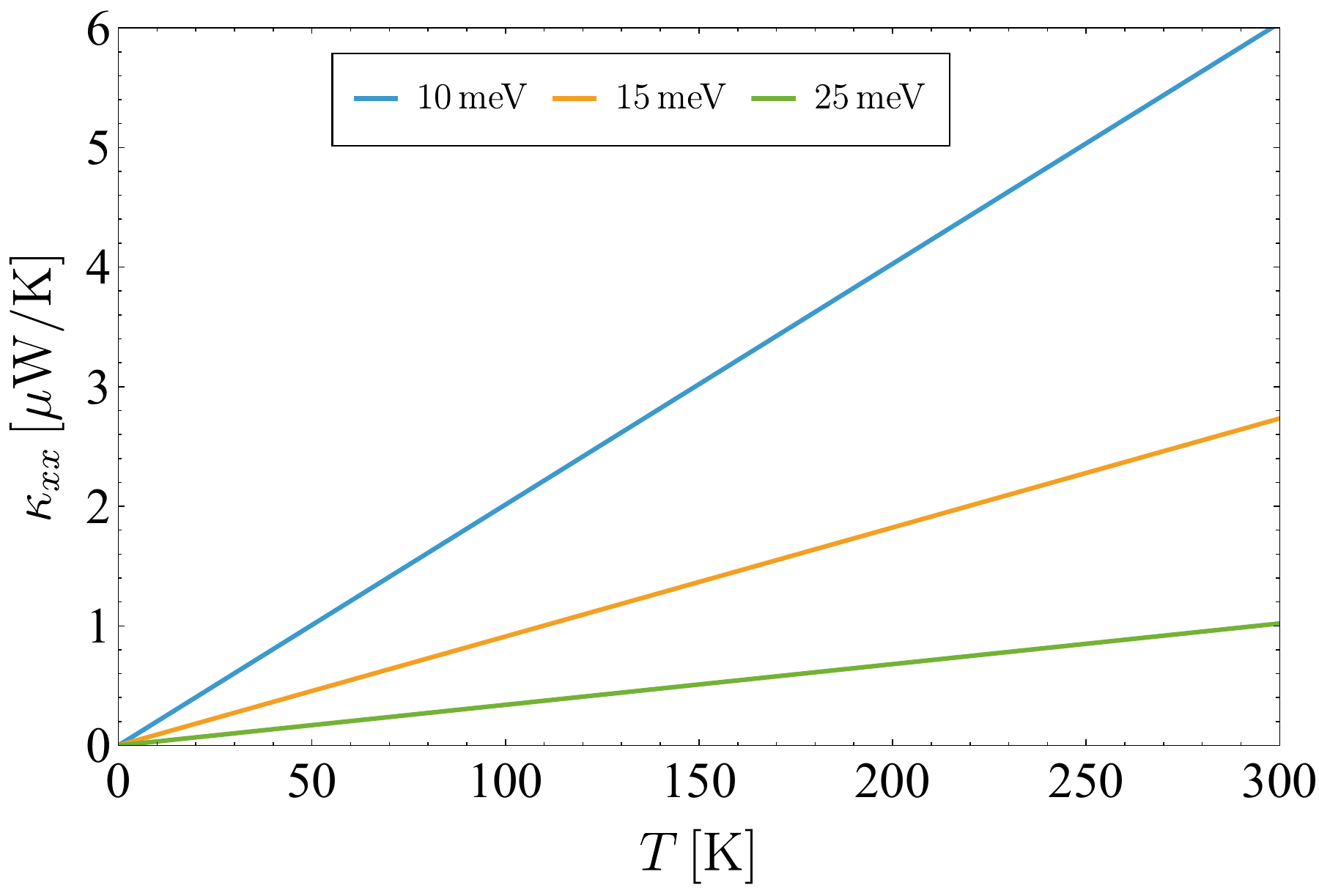}
         \subcaption{}\label{fig:cond_term_vs_T}
     \end{subfigure}
      \begin{subfigure}[]{0.45\textwidth}
         \centering
         \includegraphics[width=\linewidth]{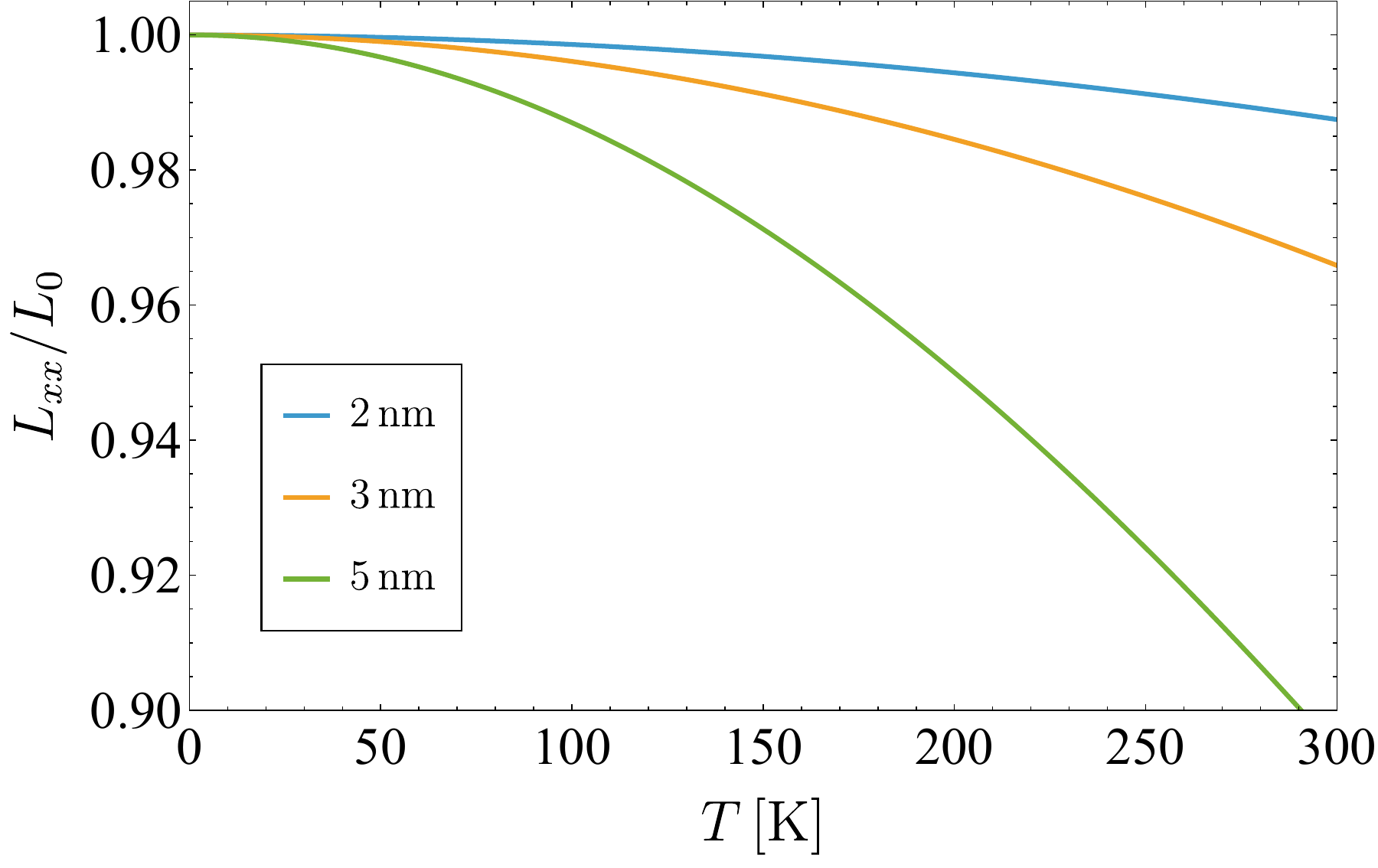}
         \subcaption{}\label{fig:lorenz_vs_T}
     \end{subfigure}
     \caption{Temperature dependence of: (a) the thermal conductivity setting $R=3$ nm, and (b) the Lorenz number setting $V_0=15$ meV. The graphs were computed for an impurity density of $n_{\mathrm{imp}} = 1 \times 10^{12}$ cm$^{-2}$, and a Fermi energy of $200$ meV.} 
    \label{fig:thermal}
     \end{figure}     
    Fig.~\ref{fig:cond_term_vs_T} shows the thermal conductivity as a function of temperature, calculated from Eq.~\eqref{eq:cond_term_final}. The computed values, on the order of $\sim$6 $\mu$W/K, are significantly lower than the total thermal conductivity values reported for monolayer graphene, which range from $\sim$1.3 MW/K (first-principles calculations \cite{Han}) to $\sim$5 MW/K (experimental measurements \cite{Balandin}). This discrepancy of several orders of magnitude is expected and highlights a key point of our study: the reported literature values are dominated by the phononic contribution (lattice thermal conductivity), whereas our calculation isolates the electronic contribution limited solely by scattering from soft-sphere impurities. Our model intentionally neglects other mechanisms, such as electron-phonon and phonon-phonon scattering, which are the primary drivers of the high thermal conductivity in pristine graphene. It is important to emphasize that the incorporation of these additional scattering mechanisms could be addressed in future work through Matthiessen’s rule \cite{ziman}. Furthermore, the presence of impurities like those studied here is known to substantially diminish the phononic thermal transport \cite{HASE2023106356}, suggesting that in real systems with defects, the electronic contribution we calculate could represent a more substantial fraction of the total thermal conductivity.
    
    \begin{figure}[!ht]
    \centering
     \begin{subfigure}[]{0.45\textwidth}
         \centering
         \includegraphics[width=\linewidth]{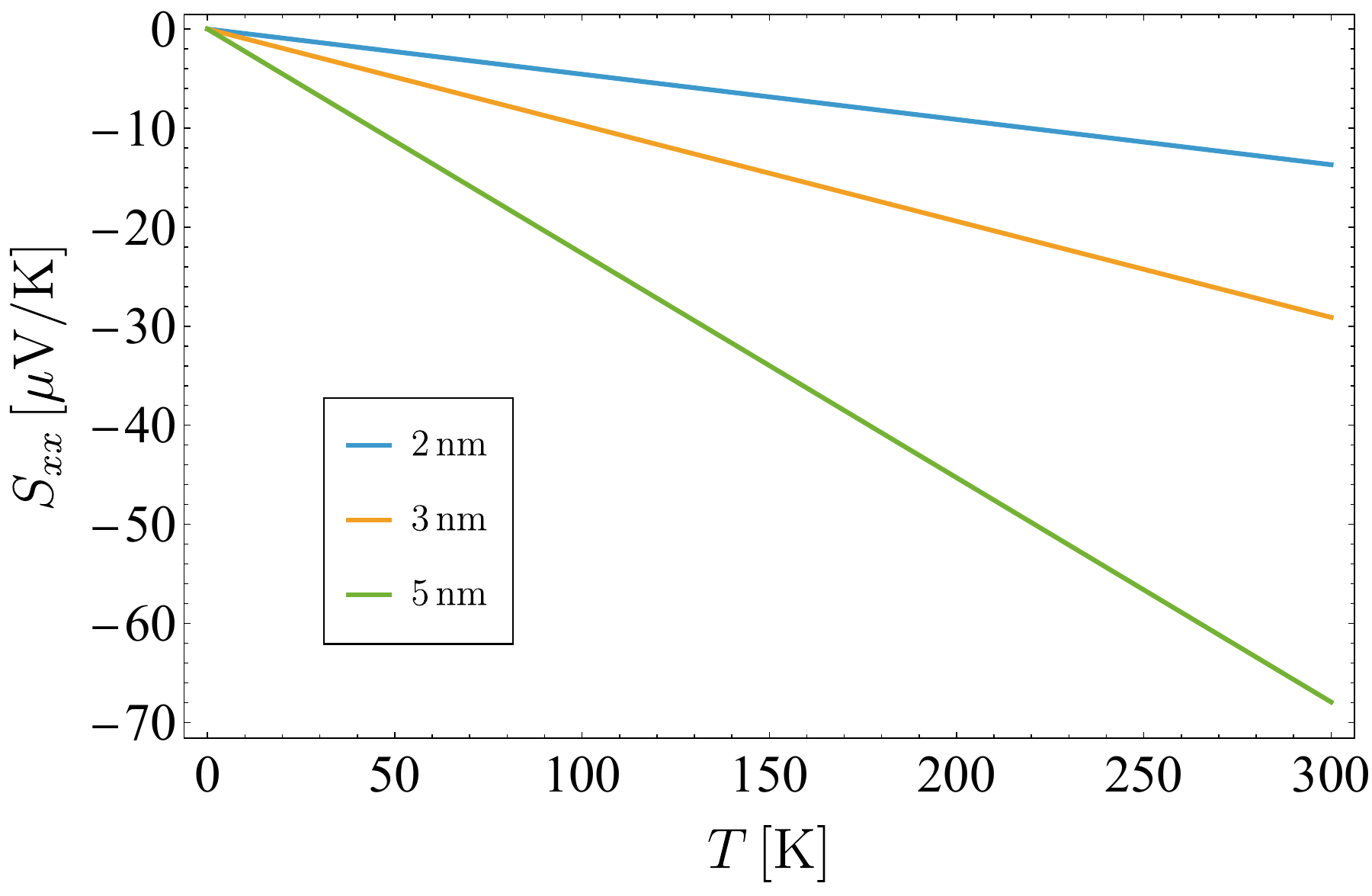}
         \subcaption{}\label{fig:Seebeck_vs_T}
     \end{subfigure}
     \begin{subfigure}[]{0.45\textwidth}
         \centering
         \includegraphics[width=\linewidth]{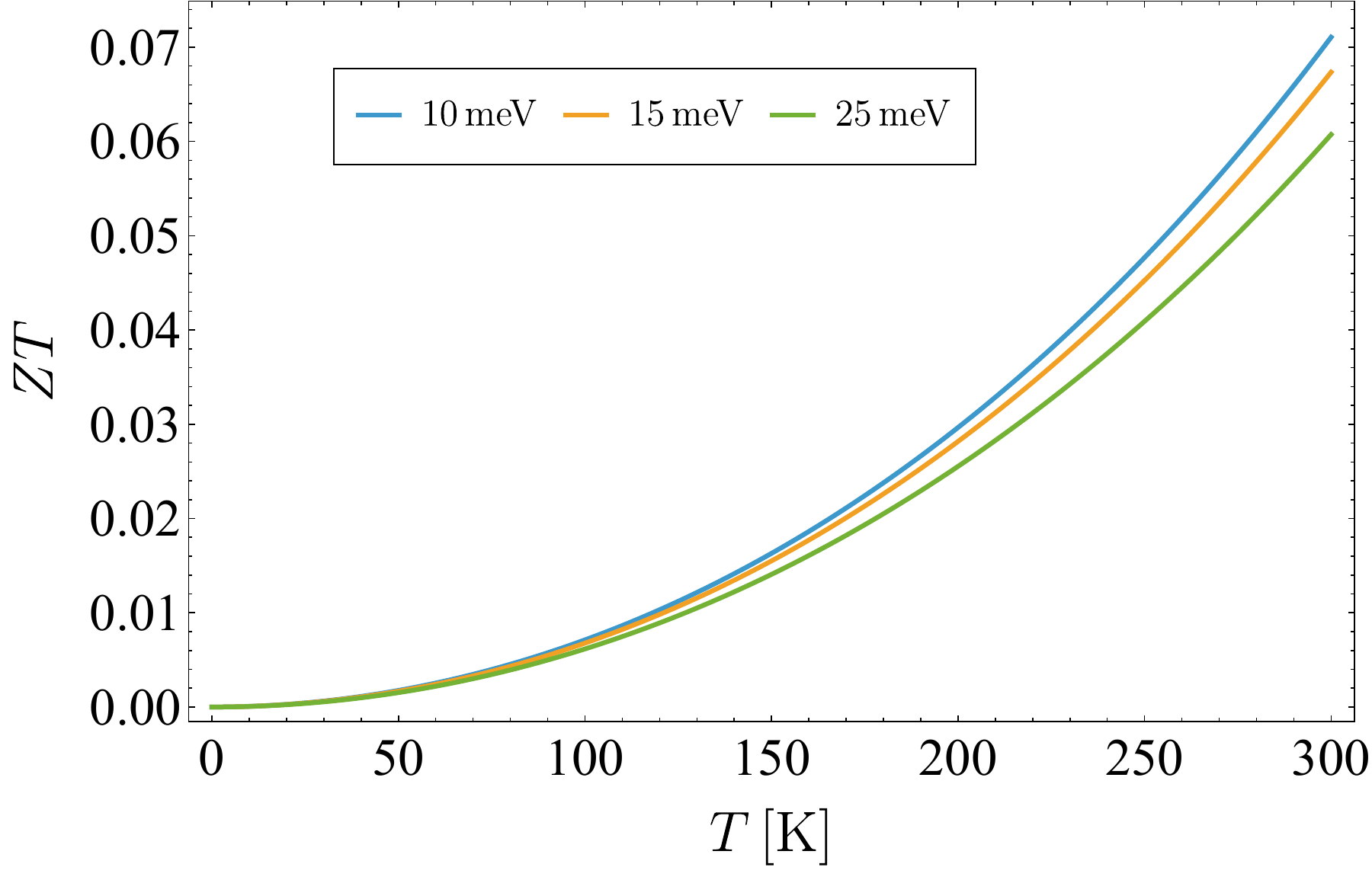}
         \subcaption{}\label{fig:ZT_vs_T}
     \end{subfigure}
     \begin{subfigure}[]{0.45\textwidth}
         \centering
         \includegraphics[width=\linewidth]{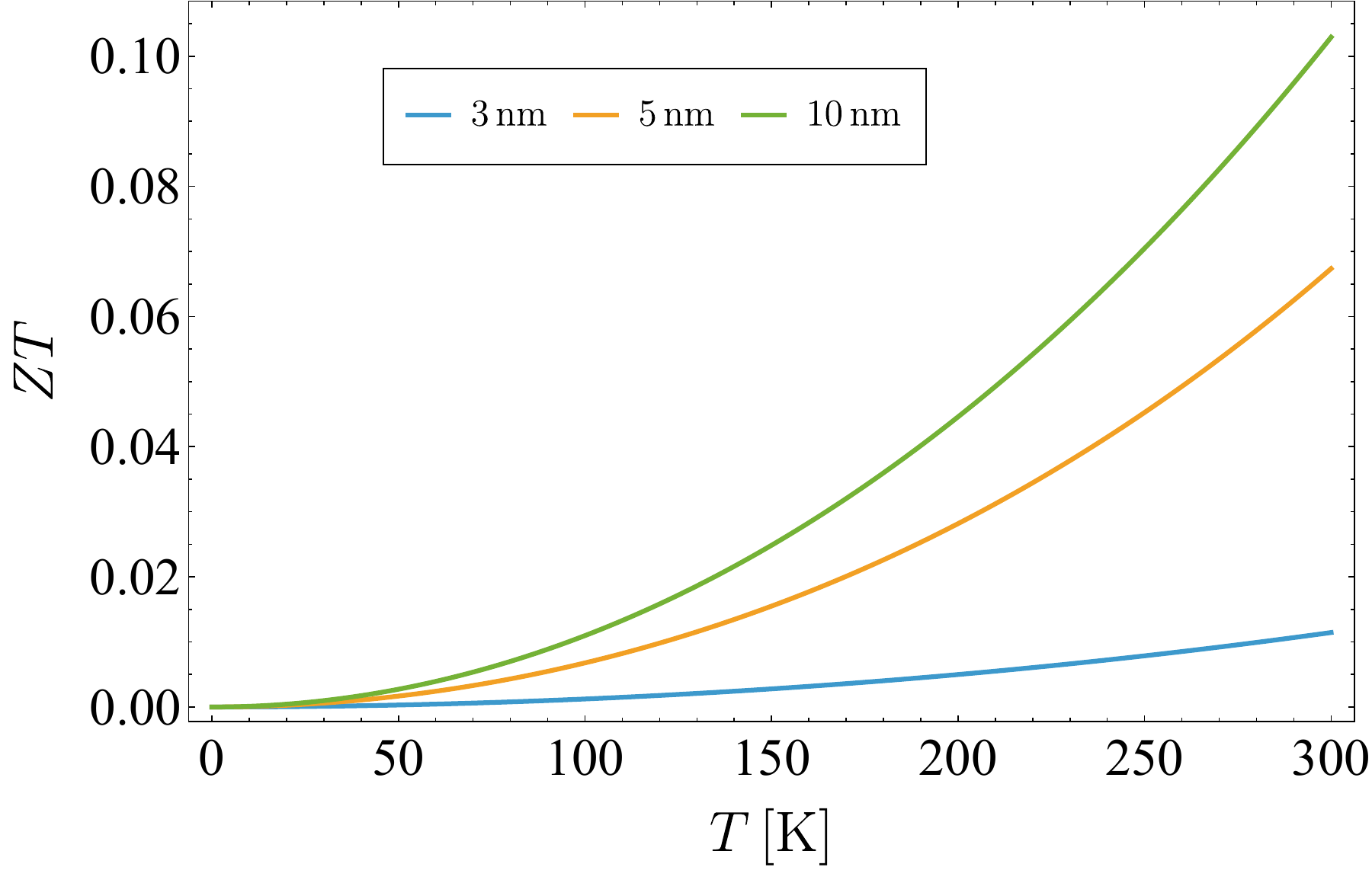}
         \subcaption{}\label{fig:ZT_vs_T-R}
     \end{subfigure}
    \caption{Temperature dependence of: (a) the Seebeck coefficient for scattering potential $V_0=15 nm$, and the figure of merit $ZT$ setting (b) $R=5$ nm, and (c) $V_0=15$ meV.  The graphs were computed for a Fermi energy of 200 meV.} 
    \label{fig:Seebeck_ZT}
    \end{figure}
    In Fig.~\ref{fig:lorenz_vs_T}, it is shown the ratio $L/L_0$, with the Lorenz number $L$ is calculated from Eq. \eqref{eq:LNumber}, plotted as a function of temperature. The results confirm the validity of the Wiedemann-Franz law in the low-temperature limit, as expected. This agreement stems from the fact that the electronic thermal conductivity in Eq.~\eqref{eq:cond_term_final} is proportional to the residual electrical conductivity given by Eq.~\eqref{eq:cond_DC_zero}. However, a clear deviation from ideal metallic behavior ($L/L_0 \neq 1$) emerges with increasing temperature. Notably, the magnitude of this deviation is highly sensitive to the radius of the scatterers; larger spheres induce a more pronounced violation of the Wiedemann-Franz law. Conversely, smaller spheres have a minimal impact, preserving the ideal metallic ratio across a wider temperature range. Furthermore, although not shown, the strength of the potential $V_0$ (varied from 5 to 50 meV) has a negligible effect on $L/L_0$, indicating that the spatial extent of the scatterer  is the crucial parameter governing the violation of the Wiedemann-Franz law in this system.

    It is instructive to place our results in the context of previous experimental and theoretical studies of the Wiedemann--Franz law in graphene. Experimentally, measurements close to the charge neutrality point have reported a strong violation of the Wiedemann--Franz law, characterized by a pronounced enhancement of the Lorenz number near the Dirac point, with the effect becoming more prominent in cleaner samples where momentum-relaxing scattering is strongly suppressed \cite{Crossno2016}. These observations have been interpreted as a signature of transport in a regime dominated by electron--hole coexistence and collective hydrodynamic effects. Tu and Das Sarma \cite{PhysRevB.107.085401,PhysRevB.108.245415} have shown that within a Boltzmann-based framework incorporating bipolar diffusion, such a maximum of the Lorenz number naturally emerges at or very close to charge neutrality. This result was further discussed in Ref. \cite{ma14112704}, where the Wiedemann–Franz law was re-derived for the case of massless Dirac fermions. In contrast, the present work focuses on transport at finite Fermi energy, away from the charge neutrality point, where a single carrier type dominates and transport is controlled by elastic scattering from smooth finite-range disorder. In this regime, hydrodynamic effects associated with electron--hole symmetry are strongly reduced, and we do not find a characteristic maximum of the Lorenz number at the Dirac point. The deviations from the Sommerfeld value obtained here instead reflect the energy dependence of scattering processes at finite doping, and are therefore consistent with the different physical regime under consideration.

\paragraph{Thermoelectric analysis.}
Figure~\ref{fig:Seebeck_vs_T} presents the calculated Seebeck coefficient, $S$, as a function of temperature, as obtained from Eq.~\eqref{eq:Seebeck}. The negative sign of $S$ is consistent with electrons being the considered charge carriers. As seen from Fig.~\ref{fig:Seebeck_vs_T}, the magnitude of the Seebeck coefficient increases with temperature, as more thermal energy is available to drive the charge carriers. A key finding is that the strength of the scattering potential, $V_0$ (varied from 5 to 50 meV), has a nearly negligible impact on the Seebeck coefficient. In contrast, the radius of the soft-sphere scatterers emerges as the central parameter governing the magnitude and functional form of $S(T)$. The computed values of $S$ reach up to $\sim$ 70 $\mu$V/K, which is consistent with the order of magnitude (40–90 $\mu$V/K) reported for doped graphene-based systems in Ref. \cite{HWANG2023467}. Notably, larger scatterer radii can yield significantly higher values of the Seebeck coefficient. Remarkably, when studying graphene nanoribbons in larger temperature regimes, the Seebeck coefficient shows a more complex profile, from which the present results can be thought as the low temperature limit \cite{ApplPhysA2024a}. 

Finally, we show the figure of merit $ZT$, calculated  as a function of temperature, in Fig.~\ref{fig:ZT_vs_T} and Fig.~\ref{fig:ZT_vs_T-R}. For practical applications, $ZT \sim 1$ is typically required, while values exceeding 2 are considered excellent. Our calculations show that $ZT$ increases with temperature, a common trend in thermoelectric materials. Figure~\ref{fig:ZT_vs_T} demonstrates that the strength of the scattering potential, $V_0$, has a relatively weak influence on $ZT$, with a slight reduction in efficiency observed for higher potentials. In contrast, Fig.~\ref{fig:ZT_vs_T-R} reveals that the radius of the scatterers is the dominant parameter controlling the magnitude of $ZT$.

The maximum values obtained in our model system reach $ZT \sim 0.1$. This result is consistent with the lower range of values reported for graphene nanoribbons (typically 0.01 to 0.5) \cite{ApplPhysA2024a,ZT_1}. Although the results presented here are not a significant improvement in thermoelectric efficiency, it presents a clear pathway for improvement: since the scatterers studied here primarily suppress electronic transport, strategies that target the reduction of $\kappa_{ph}$, such as the chevron-type geometry and isotope engineering shown to achieve $ZT \sim 2.45$ in graphene \cite{ZT_2}, could be synergistically combined with impurity engineering to achieve high efficiency. Thus, our work identifies the spatial extent of defects as a key tuning parameter for the electronic contribution to the thermoelectric performance of graphene.

\section{Conclusions} \label{conclusion_section}
We have presented a theoretical investigation of electronic and thermoelectric transport in graphene with a random distribution of impurities modeled as soft-sphere potentials. This model, which generalizes the common short-range approximation and provides a tractable proxy for smooth, screened Coulomb disorder, successfully captures the essential scattering physics of finite-range defects.

Our results demonstrate that such defects act as non-resonant scattering centers, characterized by a transport relaxation time that increases monotonically with the Fermi energy. This feature fundamentally distinguishes them from other disorder types capable of inducing quasi-bound states \cite{CBM_R2025}.

Among the most significant findings of our study, is the identification of the spatial extent (radius, $R$) of the impurities as the dominant control parameter for the thermoelectric properties of the system. While the potential strength ($V_0$) plays a secondary role, the sphere radius crucially modulates the violation of the Wiedemann-Franz law, the magnitude of the Seebeck coefficient, and ultimately, the thermoelectric figure of merit.

Although the maximum value of $ZT\sim 0.1$ obtained in our model is modest and focused on the electronic contributions, our calculations reveals a clear pathway for the optimization of graphene-based thermoelectric materials. Our analysis isolates the impurity-limited electronic contribution, demonstrating that nanoscale defect engineering (specifically, the control of defect size) is an effective strategy for enhancing the thermoelectric performance.

We therefore conclude that achieving high thermoelectric performance requires a synergistic approach. Defect engineering strategies, such as the one studied herein to maximize the electronic performance, must be combined with established methods for suppressing the dominant phononic thermal conductivity ($\kappa_{ph}$), such as nanostructuring or isotope engineering \cite{ZT_2}. This combined strategy, guided by the understanding of the dominant role of defect size, opens new avenues for the design of high-efficiency thermoelectric devices based on graphene.

\printcredits

\section*{Acknowledgments}
J.A.C. was supported by the SECIHTI PhD fellowship No. 933498. D.A.B. was supported by the DGAPA-UNAM Posdoctoral Program. J.C.P.-P was supported by the SECIHTI under the program ``Estancias Posdoctorales por México'' with CVU
number 671687.  A.M.-R. acknowledges financial support by UNAM-PAPIIT project No. IG100224, UNAM-PAPIME project No. PE109226, by SECIHTI project No. CBF-2025-I-1862 and by the Marcos Moshinsky Foundation.

\bibliographystyle{elsarticle-num}

\bibliography{cas-refs}

@article{T_Reich_2002,
    doi = {10.1103/physrevb.66.035412},
    url = {https://doi.org/10.1103/physrevb.66.035412},
    issn = {0163-1829},
    year = {2002},
    month = {Jul},
    pages = {035412},
    title = {Tight-binding description of graphene},
    author = {S. Reich and J. Maultzsch and C. Thomsen and P. Ordejón},
    number = {3},
    volume = {66},
    journal = {Physical Review B},
    publisher = {American Physical Society (APS)}}

@article{DasSarma2011ElectronicTransport,
  author    = {S. Das Sarma and Shaffique Adam and E. H. Hwang and Enrico Rossi},
  title     = {Electronic transport in two-dimensional graphene},
  journal   = {Reviews of Modern Physics},
  volume    = {83},
  number    = {2},
  pages     = {407--470},
  year      = {2011},
  month     = {May},
  doi       = {10.1103/RevModPhys.83.407},
  url       = {https://journals.aps.org/rmp/abstract/10.1103/RevModPhys.83.407}
}

@article{Tan2007Measurement,
  author    = {Y.-W. Tan and Y. Zhang and K. Bolotin and Y. Zhao and S. Adam and E. H. Hwang and S. Das Sarma and H. L. Stormer and P. Kim},
  title     = {Measurement of Scattering Rate and Minimum Conductivity in Graphene},
  journal   = {Physical Review Letters},
  volume    = {99},
  number    = {24},
  pages     = {246803},
  year      = {2007},
  month     = {December},
  doi       = {10.1103/PhysRevLett.99.246803},
  url       = {https://journals.aps.org/prl/abstract/10.1103/PhysRevLett.99.246803}
}

@article{Kim2016ValleySymmetry,
  author    = {Minsoo Kim and Ji-Hae Choi and Sang-Hoon Lee and Kenji Watanabe and Takashi Taniguchi and Seung-Hoon Jhi and Hu-Jong Lee},
  title     = {Valley-symmetry-preserved transport in ballistic graphene with gate-defined carrier guiding},
  journal   = {Nature Physics},
  volume    = {12},
  number    = {11},
  pages     = {1022--1026},
  year      = {2016},
  month     = {November},
  doi       = {10.1038/nphys3804},
  url       = {https://www.nature.com/articles/nphys3804}
}

@book{ziman_principles, address={Cambridge}, edition={2nd}, title={Principles of the Theory of Solids}, publisher={Cambridge University Press}, author={Ziman, J. M.}, year={1972}}

@book{ziman, 
address={London}, 
title={Electrons and Phonons: The Theory of Transport Phenomena}, 
publisher={Oxford, Clanderon Press}, 
author={Ziman,  J. M.}, 
year={1960}
}

@article{S_Ramezani_2011,
  title         = {
    Scattering of Dirac electrons by circular mass barriers: Valley filter and
    resonant scattering
  },
  author        = {M. Ramezani Masir and A. Matulis and F. M. Peeters},
  year          = {2011},
  month         = {Dec},
  journal       = {Physical Review B},
  publisher     = {American Physical Society (APS)},
  volume        = {84},
  number        = {24},
  pages         = {245413},
  doi           = {10.1103/physrevb.84.245413},
  issn          = {1098-0121},
  url           = {https://doi.org/10.1103/physrevb.84.245413}
}

@book{sakurai1994modern,
  title     = {Modern Quantum Mechanics},
  author    = {Sakurai, J. J.},
  year      = {1994},
  edition   = {Revised},
  publisher = {Addison-Wesley},
  isbn      = {9780201539295}
}

@article{Balandin,
	author = {Balandin, Alexander A. and Ghosh, Suchismita and Bao, Wenzhong and Calizo, Irene and Teweldebrhan, Desalegne and Miao, Feng and Lau, Chun Ning},
	booktitle = {Nano Letters},
	da = {2008/03/01},
	date = {2008/03/01},
	doi = {10.1021/nl0731872},
	isbn = {1530-6984},
	journal = {Nano Letters},
	journal1 = {Nano Lett.},
	m3 = {doi: 10.1021/nl0731872},
	month = {03},
	number = {3},
	pages = {902--907},
	publisher = {American Chemical Society},
	title = {Superior Thermal Conductivity of Single-Layer Graphene},
	ty = {JOUR},
	url = {https://doi.org/10.1021/nl0731872},
	volume = {8},
	year = {2008},
	year1 = {2008}}

@article{Han,
  title = {Thermal conductivity of monolayer graphene: Convergent and lower than diamond},
  author = {Han, Zherui and Ruan, Xiulin},
  journal = {Phys. Rev. B},
  volume = {108},
  issue = {12},
  pages = {L121412},
  numpages = {7},
  year = {2023},
  month = {Sep},
  publisher = {American Physical Society},
  doi = {10.1103/PhysRevB.108.L121412},
  url = {https://link.aps.org/doi/10.1103/PhysRevB.108.L121412}
}

@article{HWANG2023467,
title = {Large scale graphene thermoelectric device with high power factor using gradient doping profile},
journal = {Carbon},
volume = {201},
pages = {467-472},
year = {2023},
issn = {0008-6223},
doi = {https://doi.org/10.1016/j.carbon.2022.09.048},
url = {https://www.sciencedirect.com/science/article/pii/S0008622322007722},
author = {Hyeon Jun Hwang and So-Young Kim and Sang Kyung Lee and Byoung Hun Lee}
}

@Article{ZT_1,
author="Hatef Sadeghi and Sara Sangtarash and Colin J. Lambert",
title="Enhancing the thermoelectric figure of merit in engineered graphene nanoribbons",
journal="Beilstein Journal of Nanotechnology",
year="2015",
volume="6",
pages="1176-1182",
issn="2190-4286",
doi="10.3762/bjnano.6.119",
copyright="Sadeghi et al; licensee Beilstein-Institut",
publisher="Beilstein-Institut",
URL="https://doi.org/10.3762/bjnano.6.119",
}

@article{ZT_2,
	author = {Sevin{\c c}li, H{\^a}ldun and Sevik, Cem and {\c C}a{\u g}ın, Tahir and Cuniberti, Gianaurelio},
	da = {2013/02/06},
	doi = {10.1038/srep01228},
	id = {Sevin{\c c}li2013},
	isbn = {2045-2322},
	journal = {Scientific Reports},
	number = {1},
	pages = {1228},
	title = {A bottom-up route to enhance thermoelectric figures of merit in graphene nanoribbons},
	ty = {JOUR},
	url = {https://doi.org/10.1038/srep01228},
	volume = {3},
	year = {2013}}

@article{PhysRevB.91.035202,
  title = {Quantum transport in three-dimensional Weyl electron system in the presence of charged impurity scattering},
  author = {Ominato, Yuya and Koshino, Mikito},
  journal = {Phys. Rev. B},
  volume = {91},
  issue = {3},
  pages = {035202},
  numpages = {9},
  year = {2015},
  month = {Jan},
  publisher = {American Physical Society},
  doi = {10.1103/PhysRevB.91.035202},
  url = {https://link.aps.org/doi/10.1103/PhysRevB.91.035202}
}

@article{PhysRevB.84.235126,
  title = {Topological nodal semimetals},
  author = {Burkov, A. A. and Hook, M. D. and Balents, Leon},
  journal = {Phys. Rev. B},
  volume = {84},
  issue = {23},
  pages = {235126},
  numpages = {14},
  year = {2011},
  month = {Dec},
  publisher = {American Physical Society},
  doi = {10.1103/PhysRevB.84.235126},
  url = {https://link.aps.org/doi/10.1103/PhysRevB.84.235126}
}

@article{PhysRevLett.96.256602,
  title = {Quantum Hall Ferromagnetism in Graphene},
  author = {Nomura, Kentaro and MacDonald, Allan H.},
  journal = {Phys. Rev. Lett.},
  volume = {96},
  issue = {25},
  pages = {256602},
  numpages = {4},
  year = {2006},
  month = {Jun},
  publisher = {American Physical Society},
  doi = {10.1103/PhysRevLett.96.256602},
  url = {https://link.aps.org/doi/10.1103/PhysRevLett.96.256602}
}

@Article{Mao2016,
author={Mao, Jinhai
and Jiang, Yuhang
and Moldovan, Dean
and Li, Guohong
and Watanabe, Kenji
and Taniguchi, Takashi
and Masir, Massoud Ramezani
and Peeters, Francois M.
and Andrei, Eva Y.},
title={Realization of a tunable artificial atom at a supercritically charged vacancy in graphene},
journal={Nature Physics},
year={2016},
month={Jun},
day={01},
volume={12},
number={6},
pages={545-549},
issn={1745-2481},
doi={10.1038/nphys3665},
url={https://doi.org/10.1038/nphys3665}
}

@Article{Dean2010,
author={Dean, C. R.
and Young, A. F.
and Meric, I.
and Lee, C.
and Wang, L.
and Sorgenfrei, S.
and Watanabe, K.
and Taniguchi, T.
and Kim, P.
and Shepard, K. L.
and Hone, J.},
title={Boron nitride substrates for high-quality graphene electronics},
journal={Nature Nanotechnology},
year={2010},
month={Oct},
day={01},
volume={5},
number={10},
pages={722-726},
issn={1748-3395},
doi={10.1038/nnano.2010.172},
url={https://doi.org/10.1038/nnano.2010.172}
}

@article{HASE2023106356,
title = {Experimental study on the effect of impurity concentration on phonon and electronic transport properties of single-crystal silicon},
journal = {Results in Physics},
volume = {47},
pages = {106356},
year = {2023},
issn = {2211-3797},
doi = {https://doi.org/10.1016/j.rinp.2023.106356},
url = {https://www.sciencedirect.com/science/article/pii/S2211379723001493},
author = {Masataka Hase and Daiki Tanisawa and Oga Norimasa and Raichi Kamemura and Shugo Miyake and Masayuki Takashiri}
}

@article{CBM_R2025,
  title = {Thermoelectric transport in graphene under strain fields modeled by Dirac oscillators},
  author = {Cañas, Juan A. and Bonilla, Daniel A. and Martín-Ruiz, A.},
  journal = {Phys. Rev. B},
  pages = {--},
  year = {2025},
  month = {Sep},
  publisher = {American Physical Society},
  doi = {10.1103/3r17-kfy7},
  url = {https://link.aps.org/doi/10.1103/3r17-kfy7}
}

@article{DasSarmaRMP2011,
  title   = {Electronic transport in two-dimensional graphene},
  author  = {Das Sarma, S. and Adam, S. and Hwang, E. H. and Rossi, E.},
  journal = {Rev. Mod. Phys.},
  volume  = {83},
  pages   = {407--470},
  year    = {2011},
  doi     = {10.1103/RevModPhys.83.407}
}

@article{PeresRMP2010,
  title   = {Colloquium: The transport properties of graphene},
  author  = {Peres, N. M. R.},
  journal = {Rev. Mod. Phys.},
  volume  = {82},
  pages   = {2673--2700},
  year    = {2010},
  doi     = {10.1103/RevModPhys.82.2673}
}

@article{Wallace1947,
  title = {The Band Theory of Graphite},
  author = {Wallace, P. R.},
  journal = {Phys. Rev.},
  volume = {71},
  pages = {622--634},
  year = {1947},
  doi = {10.1103/PhysRev.71.622}
}

@article{NovoselovScience2004,
  title   = {Electric Field Effect in Atomically Thin Carbon Films},
  author  = {Novoselov, K. S. and Geim, A. K. and Morozov, S. V. and Jiang, D. and Zhang, Y. and Dubonos, S. V. and Grigorieva, I. V. and Firsov, A. A.},
  journal = {Science},
  volume  = {306},
  number  = {5696},
  pages   = {666--669},
  year    = {2004},
  doi     = {10.1126/science.1102896}
}

@article{NovoselovNature2005,
  title   = {Two-dimensional gas of massless Dirac fermions in graphene},
  author  = {Novoselov, K. S. and Geim, A. K. and Morozov, S. V. and Jiang, D. and Katsnelson, M. I. and Grigorieva, I. V. and Dubonos, S. V. and Firsov, A. A.},
  journal = {Nature},
  volume  = {438},
  number  = {7065},
  pages   = {197--200},
  year    = {2005},
  doi     = {10.1038/nature04233}
}

@article{CastroNetoRMP2009,
  title   = {The electronic properties of graphene},
  author  = {Castro Neto, A. H. and Guinea, F. and Peres, N. M. R. and Novoselov, K. S. and Geim, A. K.},
  journal = {Reviews of Modern Physics},
  volume  = {81},
  number  = {1},
  pages   = {109--162},
  year    = {2009},
  doi     = {10.1103/RevModPhys.81.109}
}

@article{BalandinNatMat2011,
  title   = {Thermal properties of graphene and nanostructured carbon materials},
  author  = {Balandin, Alexander A.},
  journal = {Nature Materials},
  volume  = {10},
  number  = {8},
  pages   = {569--581},
  year    = {2011},
  doi     = {10.1038/nmat3064}
}

@article{ZuevPRL2009,
  title   = {Thermoelectric and Nernst Effects in Graphene},
  author  = {Zuev, Y. M. and Chang, W. and Kim, P.},
  journal = {Physical Review Letters},
  volume  = {102},
  number  = {9},
  pages   = {096807},
  year    = {2009},
  doi     = {10.1103/PhysRevLett.102.096807}
}

@article{WeiPRL2009,
  title   = {Anomalous Thermoelectric Transport in Graphene},
  author  = {Wei, P. and Bao, W. and Pu, Y. and Lau, C. N. and Shi, J.},
  journal = {Physical Review Letters},
  volume  = {102},
  number  = {16},
  pages   = {166808},
  year    = {2009},
  doi     = {10.1103/PhysRevLett.102.166808}
}

@article{AdamPNAS2007,
  title   = {A self-consistent theory for graphene transport},
  author  = {Adam, S. and Hwang, E. H. and Galitski, V. M. and Das Sarma, S.},
  journal = {Proceedings of the National Academy of Sciences},
  volume  = {104},
  number  = {47},
  pages   = {18392--18397},
  year    = {2007},
  doi     = {10.1073/pnas.0704772104}
}

@article{HwangPRL2007,
  title   = {Carrier Transport in Two-Dimensional Graphene Layers},
  author  = {Hwang, E. H. and Adam, S. and Das Sarma, S.},
  journal = {Physical Review Letters},
  volume  = {98},
  number  = {18},
  pages   = {186806},
  year    = {2007},
  doi     = {10.1103/PhysRevLett.98.186806}
}

@article{MartinNatPhys2008,
  title   = {Observation of electron--hole puddles in graphene using a scanning single-electron transistor},
  author  = {Martin, J. and Akerman, N. and Ulbricht, G. and Lohmann, T. and Smet, J. H. and von Klitzing, K. and Yacoby, A.},
  journal = {Nature Physics},
  volume  = {4},
  number  = {2},
  pages   = {144--148},
  year    = {2008},
  doi     = {10.1038/nphys781}
}

@article{ChenNatPhys2008,
  title   = {Charged-impurity scattering in graphene},
  author  = {Chen, J. H. and Jang, C. and Adam, S. and Fuhrer, M. S. and Williams, E. D. and Ishigami, M.},
  journal = {Nature Physics},
  volume  = {4},
  number  = {5},
  pages   = {377--381},
  year    = {2008},
  doi     = {10.1038/nphys935}
}

@article{NovikovPRB2007,
  title   = {Elastic scattering theory and transport in graphene},
  author  = {Novikov, D. S.},
  journal = {Physical Review B},
  volume  = {76},
  number  = {24},
  pages   = {245435},
  year    = {2007},
  doi     = {10.1103/PhysRevB.76.245435}
}

@article{ShonAndoJPSJ1998,
author = {Shon ,Nguyen Hong and Ando ,Tsuneya},
title = {Quantum Transport in Two-Dimensional Graphite System},
journal = {Journal of the Physical Society of Japan},
volume = {67},
number = {7},
pages = {2421-2429},
year = {1998},
doi = {10.1143/JPSJ.67.2421},
URL = {https://doi.org/10.1143/JPSJ.67.2421},
eprint = {https://doi.org/10.1143/JPSJ.67.2421}
}

@article{AleinerEfetovPRL2006,
  title   = {Effect of Disorder on Transport in Graphene},
  author  = {Aleiner, I. L. and Efetov, K. B.},
  journal = {Physical Review Letters},
  volume  = {97},
  number  = {23},
  pages   = {236801},
  year    = {2006},
  doi     = {10.1103/PhysRevLett.97.236801}
}

@article{OstrovskyPRB2006,
  title   = {Electron transport in disordered graphene},
  author  = {Ostrovsky, P. M. and Gornyi, I. V. and Mirlin, A. D.},
  journal = {Physical Review B},
  volume  = {74},
  number  = {23},
  pages   = {235443},
  year    = {2006},
  doi     = {10.1103/PhysRevB.74.235443}
}

@article{RobinsonPRL2008,
  title   = {Adsorbate-Limited Conductivity of Graphene},
  author  = {Robinson, J. P. and Schomerus, H. and Oroszl{\'a}ny, L. and Fal'ko, V. I.},
  journal = {Physical Review Letters},
  volume  = {101},
  number  = {19},
  pages   = {196803},
  year    = {2008},
  doi     = {10.1103/PhysRevLett.101.196803}
}

@article{WehlingPRL2010,
  title   = {Resonant Scattering by Realistic Impurities in Graphene},
  author  = {Wehling, T. O. and Yuan, S. and Lichtenstein, A. I. and Geim, A. K. and Katsnelson, M. I.},
  journal = {Physical Review Letters},
  volume  = {105},
  number  = {5},
  pages   = {056802},
  year    = {2010},
  doi     = {10.1103/PhysRevLett.105.056802}
}

@article{FerreiraPRB2011,
  title   = {Unified description of the dc conductivity of monolayer and bilayer graphene at finite densities based on resonant scatterers},
  author  = {Ferreira, A. and Viana-Gomes, J. and Nilsson, J. and Mucciolo, E. R. and Peres, N. M. R. and Castro Neto, A. H.},
  journal = {Physical Review B},
  volume  = {83},
  number  = {16},
  pages   = {165402},
  year    = {2011},
  doi     = {10.1103/PhysRevB.83.165402}
}

@article{WuPRB2014,
  title   = {Scattering of two-dimensional massless Dirac electrons by a circular potential barrier},
  author  = {Wu, Jhih-Sheng and Fogler, Michael M.},
  journal = {Physical Review B},
  volume  = {90},
  number  = {23},
  pages   = {235402},
  year    = {2014},
  doi     = {10.1103/PhysRevB.90.235402}
}

@article{CaridadNatComm2016,
  title   = {An electrical analogy to Mie scattering},
  author  = {Caridad, Jos{\'e} M. and Connaughton, S. and Ott, C. and Weber, H. B. and Krsti{\'c}, V.},
  journal = {Nature Communications},
  volume  = {7},
  pages   = {12894},
  year    = {2016},
  doi     = {10.1038/ncomms12894}
}

@article{KlosPRB2010,
  title = {Effect of short- and long-range scattering on the conductivity of graphene: Boltzmann approach vs tight-binding calculations},
  author = {K\l{}os, J. W. and Zozoulenko, I. V.},
  journal = {Phys. Rev. B},
  volume = {82},
  issue = {8},
  pages = {081414},
  numpages = {4},
  year = {2010},
  month = {Aug},
  publisher = {American Physical Society},
  doi = {10.1103/PhysRevB.82.081414},
  url = {https://link.aps.org/doi/10.1103/PhysRevB.82.081414}
}

@article{AdamPE2008,
title = {Scattering mechanisms and Boltzmann transport in graphene},
journal = {Physica E: Low-dimensional Systems and Nanostructures},
volume = {40},
number = {5},
pages = {1022-1025},
year = {2008},
issn = {1386-9477},
doi = {https://doi.org/10.1016/j.physe.2007.09.064},
url = {https://www.sciencedirect.com/science/article/pii/S1386947707005085},
author = {Shaffique Adam and E.H. Hwang and S. {Das Sarma}}
}

@Article{Ferry2013,
author={Ferry, D. K.},
title={Short-range potential scattering and its effect on graphene mobility},
journal={Journal of Computational Electronics},
year={2013},
month={Jun},
day={01},
volume={12},
number={2},
pages={76-84},
issn={1572-8137},
doi={10.1007/s10825-012-0431-x},
url={https://doi.org/10.1007/s10825-012-0431-x}
}

@Article{Koepke2013,
author={Koepke, Justin C.
and Wood, Joshua D.
and Estrada, David
and Ong, Zhun-Yong
and He, Kevin T.
and Pop, Eric
and Lyding, Joseph W.},
title={Atomic-Scale Evidence for Potential Barriers and Strong Carrier Scattering at Graphene Grain Boundaries: A Scanning Tunneling Microscopy Study},
journal={ACS Nano},
year={2013},
month={Jan},
day={22},
publisher={American Chemical Society},
volume={7},
number={1},
pages={75-86},
issn={1936-0851},
doi={10.1021/nn302064p},
url={https://doi.org/10.1021/nn302064p}
}

@article{McCannPRL2006,
  title = {Asymmetry gap in the electronic band structure of bilayer graphene},
  author = {McCann, Edward},
  journal = {Phys. Rev. Lett.},
  volume = {96},
  pages = {086805},
  year = {2006},
  doi = {10.1103/PhysRevLett.96.086805}
}

@article{CaoNature2018SC,
  title = {Unconventional superconductivity in magic-angle graphene superlattices},
  author = {Cao, Yuan and Fatemi, Valla and Fang, Shiang and Watanabe, Kenji and Taniguchi, Takashi and Kaxiras, Efthimios and Jarillo-Herrero, Pablo},
  journal = {Nature},
  volume = {556},
  pages = {43--50},
  year = {2018},
  doi = {10.1038/nature26160}
}

@article{CaoNature2018Correlated,
  title = {Correlated insulator behaviour at half-filling in magic-angle graphene superlattices},
  author = {Cao, Yuan and Fatemi, Valla and Demir, S. and Fang, Shiang and Tomarken, Spencer L. and Luo, Jason Y. and Sanchez-Yamagishi, Javier D. and Watanabe, Kenji and Taniguchi, Takashi and Kaxiras, Efthimios and Jarillo-Herrero, Pablo},
  journal = {Nature},
  volume = {556},
  pages = {80--84},
  year = {2018},
  doi = {10.1038/nature26154}
}

@article{IJMPB2015,
author = {Rezania, Hamed},
title = {The effect of boron doping on the thermal conductivity of zigzag carbon nanotubes},
journal = {International Journal of Modern Physics B},
volume = {29},
number = {05},
pages = {1550025},
year = {2015},
doi = {10.1142/S0217979215500253}
}

@article{OptMat2016,
title = {The effects of impurity doping on the optical properties of biased bilayer graphene},
journal = {Optical Materials},
volume = {57},
pages = {8-13},
year = {2016},
issn = {0925-3467},
doi = {https://doi.org/10.1016/j.optmat.2016.04.005},
author = {Hamed Rezania and Mohsen Yarmohammadi}
}

@article{EPJB2015,
	author = {Rezania, Hamed and Abdi, Ameneh},
	date = {2015/07/06},
	doi = {10.1140/epjb/e2015-60133-3},
	id = {Rezania2015},
	isbn = {1434-6036},
	journal = {The European Physical Journal B},
	number = {7},
	pages = {173},
	title = {Thermal conductivity of disordered AA-stacked bilayer graphene in the presence of bias voltage},
	volume = {88},
	year = {2015}
}

@article{OQE2024,
	author = {Rezania, H. and Kakavandi, T.},
	date = {2024/04/25},
	doi = {10.1007/s11082-024-06696-x},
	id = {Rezania2024},
	isbn = {1572-817X},
	journal = {Optical and Quantum Electronics},
	number = {6},
	pages = {982},
	title = {Optical absorption rate in doped armchair graphene nanoribbon due to impurity atoms effects},
	volume = {56},
	year = {2024}
}

@article{ApplPhysA2024a,
	author = {Kakavandi, T. and Rezania, H.},
	date = {2024/04/23},
	doi = {10.1007/s00339-024-07512-9},
	id = {Kakavandi2024},
	isbn = {1432-0630},
	journal = {Applied Physics A},
	number = {5},
	pages = {337},
	title = {Impurity atoms effects on electronic properties and Seebeck coefficient of armchair graphene like nanoribbon},
	volume = {130},
	year = {2024}
}

@article{ApplPhysA2025,
	author = {Rezania, H. and Jamshidipour, M. and Abdi, M. and Astinchap, B.},
	date = {2025/01/09},
	doi = {10.1007/s00339-024-08231-x},
	id = {Rezania2025},
	isbn = {1432-0630},
	journal = {Applied Physics A},
	number = {2},
	pages = {94},
	title = {Thermodynamic properties of disordered kagome lattice due to spin--orbit interaction effects},
	volume = {131},
	year = {2025}
}

@article{
Crossno2016,
author = {Jesse Crossno  and Jing K. Shi  and Ke Wang  and Xiaomeng Liu  and Achim Harzheim  and Andrew Lucas  and Subir Sachdev  and Philip Kim  and Takashi Taniguchi  and Kenji Watanabe  and Thomas A. Ohki  and Kin Chung Fong },
title = {Observation of the Dirac fluid and the breakdown of the Wiedemann-Franz law in graphene},
journal = {Science},
volume = {351},
number = {6277},
pages = {1058-1061},
year = {2016},
doi = {10.1126/science.aad0343},
URL = {https://www.science.org/doi/abs/10.1126/science.aad0343},
eprint = {https://www.science.org/doi/pdf/10.1126/science.aad0343}}

@article{PhysRevB.107.085401,
  title = {Wiedemann-Franz law in graphene},
  author = {Tu, Yi-Ting and Das Sarma, Sankar},
  journal = {Phys. Rev. B},
  volume = {107},
  issue = {8},
  pages = {085401},
  numpages = {13},
  year = {2023},
  month = {Feb},
  publisher = {American Physical Society},
  doi = {10.1103/PhysRevB.107.085401},
  url = {https://link.aps.org/doi/10.1103/PhysRevB.107.085401}
}

@article{PhysRevB.108.245415,
  title = {Wiedemann-Franz law in graphene in the presence of a weak magnetic field},
  author = {Tu, Yi-Ting and Das Sarma, Sankar},
  journal = {Phys. Rev. B},
  volume = {108},
  issue = {24},
  pages = {245415},
  numpages = {6},
  year = {2023},
  month = {Dec},
  publisher = {American Physical Society},
  doi = {10.1103/PhysRevB.108.245415},
  url = {https://link.aps.org/doi/10.1103/PhysRevB.108.245415}
}

@Article{ma14112704,
AUTHOR = {Rycerz, Adam},
TITLE = {Wiedemann–Franz Law for Massless Dirac Fermions with Implications for Graphene},
JOURNAL = {Materials},
VOLUME = {14},
YEAR = {2021},
NUMBER = {11},
ARTICLE-NUMBER = {2704},
URL = {https://www.mdpi.com/1996-1944/14/11/2704},
PubMedID = {34063902},
ISSN = {1996-1944},
DOI = {10.3390/ma14112704}
}

@article{PhysRevB.91.115410,
  title = {Thermoelectric effect enhanced by resonant states in graphene},
  author = {Inglot, M. and Dyrda\l{}, A. and Dugaev, V. K. and Barna\ifmmode \acute{s}\else \'{s}\fi{}, J.},
  journal = {Phys. Rev. B},
  volume = {91},
  issue = {11},
  pages = {115410},
  numpages = {7},
  year = {2015},
  month = {Mar},
  publisher = {American Physical Society},
  doi = {10.1103/PhysRevB.91.115410},
  url = {https://link.aps.org/doi/10.1103/PhysRevB.91.115410}
}



\end{document}